\documentclass[journal]{IEEEtran}

\usepackage[numbers, sort&compress]{natbib}

\usepackage{amsmath, amsfonts, latexsym, siunitx}
\usepackage{amssymb}
\usepackage{bm}

\usepackage{empheq}
\usepackage{booktabs}
\usepackage{threeparttable}
\usepackage{multirow}
\usepackage{diagbox}

\usepackage[linesnumbered, ruled]{algorithm2e}

\usepackage{graphicx}
\usepackage{subfigure}

\usepackage[english]{babel}

\usepackage{bbm}

\usepackage{enumerate}
\usepackage{paralist}

\usepackage{arydshln}

\usepackage{url}
\usepackage{hyperref}

\usepackage{color}

\begin{document}

\renewcommand{\figurename}{Fig.}
\renewcommand{\tablename}{TABLE}


\title{Radio Generation Using Generative \\ Adversarial Networks with An Unrolled Design}


\author{Weidong~Wang, Jiancheng~An, Hongshu~Liao, Lu~Gan, and Chau Yuen, \IEEEmembership{Fellow, IEEE}}

\markboth{IEEE Transactions on Cognitive Communications and Networking}
{Wang \MakeLowercase{\textit{et al.}}: Generative Adversarial Networks}

\maketitle

\begin{abstract}
    As a revolutionary generative paradigm of deep learning, generative adversarial networks (GANs) have been widely applied in various fields to synthesize realistic data. However, it is challenging for conventional GANs to synthesize raw signal data, especially in some complex cases. In this paper, we develop a novel GAN framework for radio generation called ``Radio GAN". Compared to conventional methods, it benefits from three key improvements. The first is learning based on sampling points, which aims to model an underlying sampling distribution of radio signals. The second is an unrolled generator design, combined with an estimated pure signal distribution as a prior, which can greatly reduce learning difficulty and effectively improve learning precision. Finally, we present an energy-constrained optimization algorithm to achieve better training stability and convergence. Experimental results with extensive simulations demonstrate that our proposed GAN framework can effectively learn transmitter characteristics and various channel effects, thus accurately modeling for an underlying sampling distribution to synthesize radio signals of high quality.
\end{abstract}


\IEEEpeerreviewmaketitle

\section{Introduction} \label{Section: Introduction}
\IEEEPARstart{G}{enerative} adversarial networks (GANs) \cite{goodfellow2014generative} are a popular generative paradigm of deep learning. The ability to effectively model complex real-world data makes GANs widely used in various domains \cite{hong2019generative}. The most significant impact has occurred in computer vision \cite{wang2021generative}, where great advances have been made in challenges such as plausible image generation, including those more fine-grained tasks, e.g., image-to-image translation \cite{liu2017unsupervised, choi2018stargan, ma2018gan} and image super-resolution \cite{ledig2017photo, wang2018esrgan}. In this paper, we will explore using GANs to synthesize wireless communication signal data, known as radio generation. Previous literature has shown many efforts to exploit GANs for radio generation. But they generally focus on specific tasks, among which most are related to radio recognition, including modulation recognition and radio frequency (RF) fingerprinting, instead of radio generation itself.

Earlier, in modulation recognition, Tang \emph{et al.} \cite{tang2018digital} adopted an auxiliary classifier GAN (ACGAN) to enlarge their dataset, i.e., data augmentation, which can improve supervised performance when there is a lack of sufficient labeled samples. Notably, they use an image-like representation for signal data. Similarly, Patel \emph{et al.} \cite{patel2020data} and Lee \emph{et al.} \cite{lee2021uniqgan} employed a conditional GAN (CGAN) and a variant of ACGAN for data augmentation, respectively, but both chose to synthesize raw signal data. The conversion from raw signal data to image-like data inevitably leads to a certain loss of information that makes downstream tasks unable to achieve competitive performance. Consequently, it is usually recommended to synthesize raw signal data, typically represented in an in-phase and quadrature (I/Q) form. The authors in \cite{li2018radio, li2018generative, zhou2020generative} further developed semi-supervised solutions based on GANs to utilize a large amount of unlabeled signal data that is easily acquired. In particular, GANs can be employed to generate fake signals that cannot be reliably distinguished from intended signals for spoofing wireless signals \cite{shi2019generative, roy2019rfal}, which allows RF fingerprinting to be more robust with potentially malicious attacks.

Except for radio recognition, such radio generation can also play an effective role in many other wireless communication applications. For instance, O'Shea \emph{et al.} \cite{o2018physical} adopted a GAN to approximate responses of communication channels and employed it to find an optimal physical-layer encoding scheme. Likewise, Ye \emph{et al.} \cite{ye2020deep} employed a GAN to represent channel impairments, then used it to help train an end-to-end communication system where transmitters may utilize channel state information to optimize their transmissions. Undoubtedly, such radio generation with GANs still has great potential to be explored in wireless communication.

In general, we find that almost all previous studies have focused on specific tasks, and they generally ignore what kinds of quality their generated signal data are, although this may lack a generally accepted evaluation metric. However, such quality for generated data often essentially determines performance. Taking data augmentation in radio recognition as an example. In theory, a GAN can synthesize an infinite number of valid samples from random noise if it has successfully modeled for a sample distribution, thus improving supervised performance nearly close to fully supervised learning with an ideal number of samples. However, we observe that this performance gain is very limited in practice, which is often not as effective as other commonly used augmentation operations for signal data, e.g., rotation and flipping \cite{huang2019data}, and sometimes even worse than a baseline, especially with relatively complex cases, such as considering channel fading. To our best knowledge, this issue has never been carefully discussed in any previous literature.

In this paper, we shall systematically analyze why it is challenging for conventional GANs to synthesize raw signal data. To this end, we develop a novel GAN framework for radio generation, called ``Radio GAN", which benefits from three key improvements:
\begin{itemize}
    \item We propose to train GANs on sampling points, which aims to model an underlying sampling distribution of radio signals instead of a radio sample distribution.
    \item We present a novel generator design that unrolls a holistic generator into a combination of multiple neural network modules designed individually with domain knowledge, and in conjunction with an estimated pure signal distribution, it can greatly reduce learning difficulty and effectively increase modeling precision.
    \item We present an energy-constrained optimization algorithm to achieve better training stability and convergence.
\end{itemize}
Extensive experiments are conducted to validate our proposed GAN framework for radio generation. The experimental results demonstrate that our proposed framework can effectively learn transmitter characteristics and various channel effects, thus accurately modeling an underlying sampling distribution of radio signals. Compared to conventional GANs, ours has better modeling effects, which allows it to synthesize signal data of higher quality, and thus achieve satisfactory performance in downstream tasks.

\vspace{11pt}

The rest of this paper is organized as follows. The preliminaries of GANs are described in Section~\ref{Section: Generative Adversarial Network}. The proposed GAN framework for radio generation is depicted in Section~\ref{Section: Methodology}. The experiments and related results are given in Section~\ref{Section: Experiments and Results} before concluding in Section~\ref{Section: Conclusion}.

\section{Generative Adversarial Network} \label{Section: Generative Adversarial Network}
The technological breakthrough brought by Generative Adversarial Networks (GANs) has rapidly produced revolutionary consequences on machine learning \cite{wang2021generative, cai2021generative}. An original version of GANs (often termed ``vanilla GAN") was introduced by Goodfellow \emph{et al.} \cite{goodfellow2014generative} in 2014, and it typically consists of a generator $G$ and a discriminator $D$, which are both implemented as deep neural networks. The generator samples a latent variable $\boldsymbol{z}$ from a prior distribution $p_{\boldsymbol{z}}$ (e.g., Gaussian) as input and then maps it as data samples that resemble a target distribution $p_{\mathrm{data}}$. The discriminator is a binary classifier that learns to distinguish whether a given sample is real or generated as far as possible. The training of vanilla GANs is to solve a minimax game between $G$ and $D$, as shown below.
\begin{equation}
    \begin{split}
        & \min_{G} \max_{D} \mathcal{L}(G, \, D) = \\
        & \mathbb{E}_{\boldsymbol{x} \sim p_\mathrm{data}(\boldsymbol{x})} \left[ \log D(\boldsymbol{x}) \right] + \mathbb{E}_{\boldsymbol{z} \sim p_{\boldsymbol{z}}(\boldsymbol{z})} \left[ \log (1 - D(G(\boldsymbol{z})) \right]
    \end{split}
\end{equation}
The objective loss is equivalent to minimizing the Jensen-Shannon (JS) divergence between $p_{{\scriptscriptstyle G}(\boldsymbol{z})}$ and $p_{\mathrm{data}}$. In theory, GANs are capable of modeling an arbitrarily complex probability distribution, and Goodfellow \emph{et al.} \cite{goodfellow2014generative} have proved that there exists a unique solution at $D = 1/2$, corresponding to a global Nash equilibrium \cite{nash1951non}. In practice, however, GANs have struggled to reach this Nash equilibrium \cite{salimans2016improved}.

Such vanilla GANs suffer from many other issues, including non-convergence, mode collapse, and gradient vanishing, making their training process highly unstable, especially when modeling for relatively complex data \cite{wiatrak2019stabilizing}. Many tricks have been introduced to stabilize training, such as careful selection of network architectures \cite{radford2015unsupervised}, mini-batch discrimination \cite{salimans2016improved}, and noise injection \cite{sonderby2016amortised}. The most effective way is to exploit other alternative loss functions for optimization instead of the JS divergence, e.g., optimizing a Wasserstein distance, including WGAN \cite{arjovsky2017wasserstein} and WGAN-GP \cite{gulrajani2017improved}. The discriminator of these variants is enforced to have Lipschitz continuity by gradient regularization or normalization, learned as an energy function and no longer a binary classifier. They can be written in a general form:
\begin{equation}
    \min_G \max_{\Vert D\Vert_{L}\leq 1} \mathbb{E}_{\boldsymbol{x} \sim p_\mathrm{data}(\boldsymbol{x})}\left[D(\boldsymbol{x})\right] - \mathbb{E}_{\boldsymbol{z} \sim p_{\boldsymbol{z}}(\boldsymbol{z})}\left[D(G(\boldsymbol{z}))\right]
\end{equation}

Moreover, conditional GANs (CGANs) \cite{mirza2014conditional} are an important extension of GANs that allow us to introduce additional information, such as class labels, which can largely resolve mode collapse. Auxiliary classifier GANs (ACGANs) \cite{odena2017conditional} can corporate an additional classification loss in training. Refer to \cite{saxena2021generative, jabbar2021survey} for more comprehensive reviews about other GAN variants and recent progress.

\section{Methodology} \label{Section: Methodology}

\subsection{Mechanism Analysis}
To begin with, we shall systematically analyze why it is so challenging for GANs to synthesize raw signal data. It is important to emphasize that conventional GANs for such radio generation are typically trained on radio samples, which aims to learn a radio sample distribution. The radio sample is a long sequence that consists of many complex-valued sampling points if its baseband form is considered, each of which can be considered drawn from an underlying sampling distribution of radio signals. The sampling points in a radio sample could be mutually independent but are more likely to be partially correlated due to filtering or other similar factors.

That is, a multi-dimensional probability distribution with an independent or non-independent and identically distributed assumption needs to be modeled with GANs. The challenge comes from two aspects. First, this identically distributed nature cannot be strictly guaranteed in practice, and specifically, deep neural networks cannot effectively learn a weight matrix with each column identical due to random initialization and stochastic gradient descent (SGD) \cite{goodfellow2016deep}, especially with relatively high dimensionality. The second is that this underlying sampling distribution itself is relatively complex. The identically distributed nature, though cannot be strictly guaranteed, could be largely approximated in practice. The latter actually plays a decisive role.

\subsection{Learning on Sampling Points}
Since this identically distributed nature cannot be strictly guaranteed, we propose not to model a radio sample distribution directly but turn to learn its corresponding underlying sampling distribution only. The issue thus can be effectively circumvented. In this case, a GAN should be trained on sampling points instead of radio samples. Learning a low-dimensional probability distribution is also often easier than a higher-dimensional one. In addition, this strategy can greatly reduce demand for training data, as a single radio sample can be broken up into many sampling points, which might be very useful in some downstream tasks, e.g., data augmentation.

This strategy requires a pure signal as extra prior information to ensure symbol order. Otherwise, only a large number of sampling points without any correlation are generated but cannot comprise a coherent waveform. The ``pure" denotes a signal that has not yet suffered from any transmitter characteristics, channel effects, and other possible considerations for wireless communication, which can be implemented by communication simulation. To this end, we need to know several necessary signal parameters, including modulation scheme, oversampling ratio, rolloff factor of pulse shaping, and symbol distribution. In cooperative communications, it is usually not difficult to obtain these parameters. Even in non-cooperative communications, we can still estimate them from received signal data \cite{wang2010nda, shi2007blind, thomas2017blind}. It is worth mentioning that introducing such a pure signal also helps to learn underlying sampling distributions.

\subsection{Network Design via Model Unrolling}
Even learning on sampling points, we find that conventional GANs still cannot accurately model for an underlying sampling distribution of radio signals, especially with relatively complex cases, such as considering channel fading. The underlying sampling distribution is jointly affected by associated modulation parameters, specific transmitter characteristics, various channel effects, and other possible considerations for wireless communication, and then these factors accordingly determine its complexity. Inspired by this fact, we present a novel generator design that unrolls a holistic generator into an organic combination of multiple neural network modules designed individually with domain knowledge.

Before diving into our design, it is worth introducing $4$ kinds of wireless communication signal models. From simple to complex, we consider one, or more, or all communication elements among additive channel effects, multiplicative channel effects, and specific transmitter characteristics. If only additive channel effects, such as channel noise, are considered, we can abstract a received signal as
\begin{equation}
    x(t) = s(t) + n(t)
\end{equation}
where $s(t)$ denotes a pure signal and assumes that it obeys a known probability distribution $p_{s}$, and $n(t)$ represents such additive channel effects, subject to an unknown probability distribution $p_{n}$. Then, we can further consider multiplicative channel effects, such as channel fading, corresponding to a received signal abstracted as
\begin{equation}
    x(t) = \alpha(t)s(t) + n(t)
\end{equation}
where $\alpha(t)$ represents such multiplicative channel effects, subject to an unknown probability distribution $p_{\alpha}$. For clarity, we refer to these two wireless communication signal models as $\mathrm{Signal}_{\, n}$ and $\mathrm{Signal}_{\, \alpha, \, n}$, respectively.

Sometimes, e.g., in RF fingerprinting, we must consider transmitter characteristics. It is important to emphasize that such transmitter characteristics are not equivalent to RF fingerprints. Only those device-specific features are termed RF fingerprints. The transmitter characteristics are mainly derived from digital-to-analog (D/A) converters, local frequency synthesizers, and power amplifiers' nonlinearity, which can be synthetically represented using a non-linear mapping $h(\cdot)$. The received signal thus can be rewritten as
\begin{equation}
    x(t) = h(s(t)) + n(t)
\end{equation}
and
\begin{equation}
    x(t) = \alpha(t)h(s(t)) + n(t)
\end{equation}
For clarity, we refer to them as $\mathrm{Signal}_{\, h, \, n}$ and $\mathrm{Signal}_{\, h, \, \alpha, \, n}$, respectively. Then, we can build corresponding unrolled generators according to wireless communication signal models.

\begin{figure}[htb]
    \centering
    \includegraphics[scale=0.56]{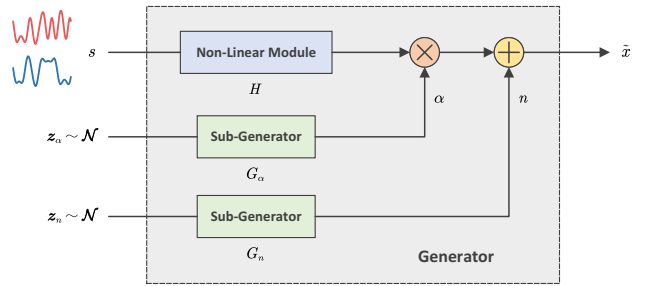}
    \caption{The unrolled generator for $\mathrm{Signal}_{\, h, \, \alpha, \, n}$.}
    \label{Figure: Generator}
\end{figure}

Taking $\mathrm{Signal}_{\, h, \, \alpha, \, n}$ as an example, we build an unrolled generator illustrated in Fig. \ref{Figure: Generator}, and it consists of a non-linear transformation module $H$, two sub-generators, $G_{\alpha}$ and $G_{n}$, as well as a multiplication node and an addition node. It is not difficult to find that introducing a pure signal distribution as a priori plays a crucial role in our design. The non-linear transformation module is used to learn transmitter characteristics, and these two sub-generators are used to learn multiplicative channel effects and additive channel effects, respectively. The specific design details are as follows.

\vspace{11pt}

\begin{figure}[htb]
    \centering
    \includegraphics[scale=0.65]{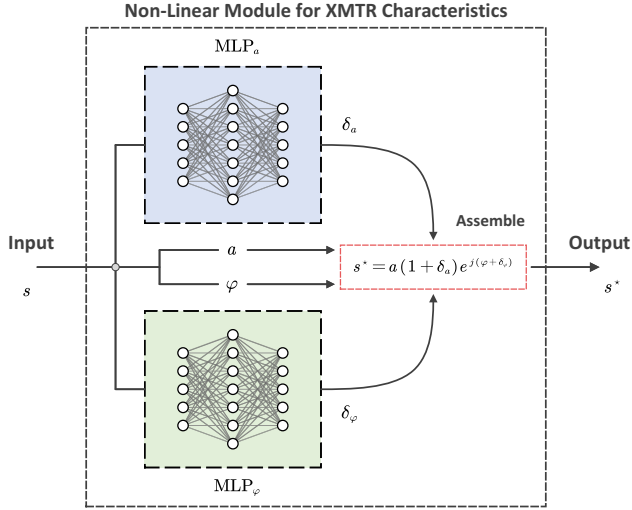}
    \caption{The design for non-linear transformation module, $H$.}
    \label{Figure: Non-Linear Transformation Module}
\end{figure}

\noindent\textbf{Non-Linear Transformation Module} As illustrated in Fig. \ref{Figure: Non-Linear Transformation Module}, we do not directly use a holistic neural network to learn such transmitter characteristics but once again take full advantage of domain knowledge to design a more refined structure. For a given complex-valued input $s$ sampled from $p_{s}$, we first compute its amplitude and phase, denoted by $\alpha$ and $\varphi$. This input is mapped as its corresponding amplitude and phase offsets, denoted by $\delta_{a}$ and $\delta_{\varphi}$, using two non-linear multilayer perceptrons (MLPs), as follows.
\begin{equation}
    \left\{ \begin{aligned}
        \delta_{a}       & =\mathrm{MLP}_a(s)         \\
        \delta_{\varphi} & =\mathrm{MLP}_{\varphi}(s) \\
    \end{aligned} \right.
\end{equation}
Then, $a$, $\varphi$, $\delta_{a}$, $\delta_{\varphi}$ are assembled as an output with specific transmitter characteristics, denoted by $s^{\star}$, according to
\begin{equation}
    s^{\star} = a(1 + \delta_{a})e^{j(\varphi +\delta_{\varphi})}
\end{equation}
If a single device could be learnable, it must also be feasible to learn multiple devices, referenced from
\begin{equation}
    s^{\star} =\left[ \begin{matrix}
            h_1(s) & h_2(s) & ... & \\
        \end{matrix} \right] \odot y
\end{equation}
which indicates that when learning for multiple devices, we must introduce device labels ($y$) since sampling points from a pure signal do not involve any transmitter information.

\begin{figure}[htb]
    \centering
    \includegraphics[scale=0.8]{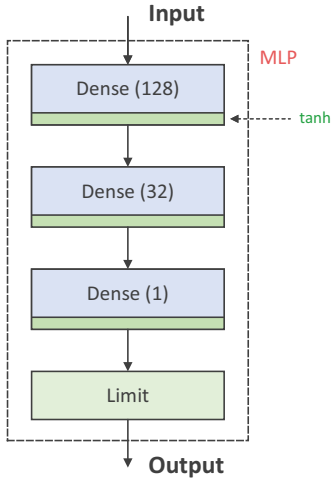}
    \caption{The structure of designed non-linear MLPs.}
    \label{Figure: MLP}
\end{figure}

The network structure is identical for $\operatorname{MLP}_{a}$ and $\operatorname{MLP}_{\varphi}$, as illustrated in Fig. \ref{Figure: MLP}, and it consists of three dense layers activated using $\tanh$ and ends with a numerical limit. These three dense layers have $128$, $32$, and $1$ neurons, respectively, which could be adjusted with specific demands in practice. The numerical limit employs a specified weight and bias to transform output from $\tanh$'s $(-1, \, 1)$ to a desired numerical interval. For example, we can usually configure such output numerical intervals as $(-0.5, \, 0.5)$ for $\operatorname{MLP}_{a}$ to learn $\delta_{a}$ and $(-\frac{\pi}{2}, \, \frac{\pi}{2})$ for $\operatorname{MLP}_{\varphi}$ to learn $\delta_{\varphi}$.

\vspace{11pt}

\noindent\textbf{Sub-Generator} The channel effects necessarily follow a real-valued distribution in reality, and it can be transformed into an equivalent complex-valued form via orthogonal transformation. If we treat this complex-valued distribution as a two-dimensional real-valued distribution, its two marginal probability density functions must be identical. The sub-generator thus only needs to learn a marginal probability distribution.

\begin{figure}[htb]
    \centering
    \includegraphics[scale=0.65]{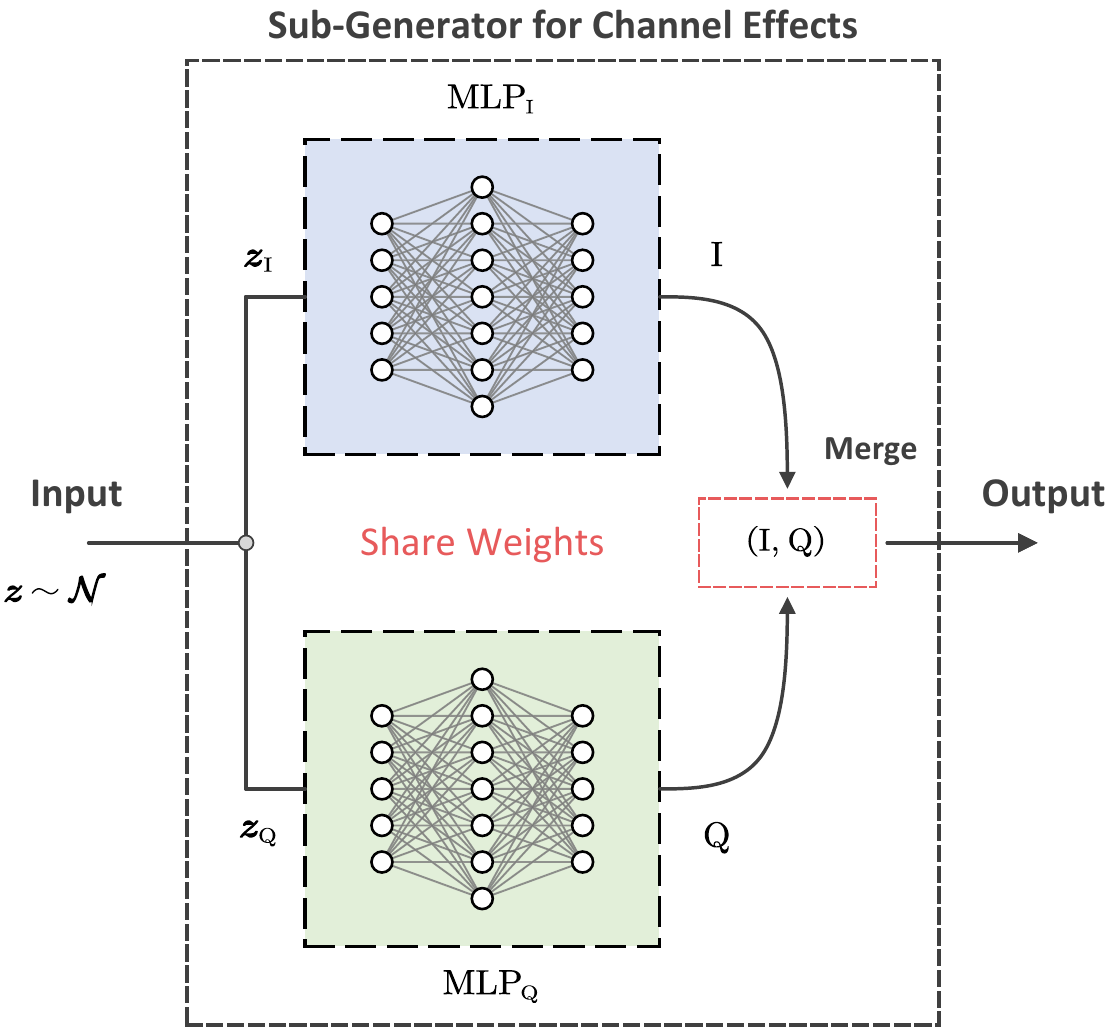}
    \caption{The design for sub-generators, $G_{\alpha}$ and $G_{n}$.}
    \label{Figure: Sub-Generator}
\end{figure}

As illustrated in Fig. \ref{Figure: Sub-Generator}, we adopt two non-linear MLPs to learn marginal probability distributions but make them share weights to ensure that the same one has been learned. This design can also ensure that each output dimension is strictly independent. The non-linear MLP used here is still consistent with Fig. \ref{Figure: MLP} but has a different input size. Each MLP will sample from an $8$-dimensional Gaussian distribution. The whole sub-generator thus samples from a $16$-dimensional Gaussian distribution. The input and network size could be appropriately increased when learning for a relatively complex distribution of channel effects.

\vspace{11pt}

\noindent\textbf{Multiplication and Addition Nodes} The generator architecture includes two arithmetic nodes. The multiplication node performs complex multiplication rather than element-wise multiplication. The addition node performs simple element-wise addition.

\vspace{11pt}

Note that both $G_{\alpha}$ and $G_{n}$ sample from a multi-dimensional Gaussian distribution, i.e., $\boldsymbol{z}_{\alpha} \sim p_{\boldsymbol{z}_{\alpha}}$ and $\boldsymbol{z}_{n} \sim p_{\boldsymbol{z}_{n}}$. For simplicity, we denote $\boldsymbol{z}_{\alpha}$ and $\boldsymbol{z}_{n}$ uniformly by $\boldsymbol{z}$. The unrolled generators corresponding to all $4$ kinds of wireless communication signal models thus can be uniformly expressed as $G_{\mathrm{unrolled}}(\boldsymbol{z} \mid s)$. If multiple classes of signal data are generated, such as considering multiple devices for RF fingerprinting, it can be expressed as $G_{\mathrm{unrolled}}(\boldsymbol{z} \mid s, \, y)$ after introducing labels to guide training. The unrolled generator design has been elaborated on.

The unrolled design is expected to improve modeling precision. This design naturally inherits a lot of prior structures from theoretical signal models rather than learning that information from intensive data. The specific network structure can inform what kind of underlying mapping a GAN should learn, effectively reducing its function search space. More considerations for wireless communication can be integrated using a similar fashion.

\subsection{Optimization Schemes}
To summarize, we have discussed $4$ kinds of wireless communication signal models:
\begin{itemize}
    \setlength{\itemsep}{6pt}
    \item $\mathrm{Signal}_{\, n}$: $x(t) = s(t) + n(t)$
    \item $\mathrm{Signal}_{\, \alpha, \, n}$: $x(t) = \alpha(t)s(t) + n(t)$
    \item $\mathrm{Signal}_{\, h, \, n}$: $x(t) = h(s(t)) + n(t)$
    \item $\mathrm{Signal}_{\, h, \, \alpha, \, n}$: $x(t) = \alpha(t)h(s(t)) + n(t)$
\end{itemize}
which correspond to different unrolled generators. For clarity, we refer to their respective GANs as $\mathrm{GAN}_{\, n}$, $\mathrm{GAN}_{\, \alpha, \, n}$, $\mathrm{GAN}_{\, h, \, n}$, and $\mathrm{GAN}_{\, h, \, \alpha, \, n}$. Then, we provide each GAN with a specialized optimization scheme, as shown below.

\vspace{11pt}

\noindent\textbf{Optimization of $\mathrm{GAN}_{\, n}$} Structurally, $\mathrm{GAN}_{\, n}$ is not much different from conventional GANs, and both have only one unknown probability distribution to be learned. Hence, we can optimize it directly with any advanced GAN training techniques. This work uniformly employs gradient normalization, i.e., GN-GAN \cite{wu2021gradient}, for training, as given in Algorithm \ref{Algorithm: GN-GAN for Radio Generation}.

As mentioned earlier, GANs can accurately model a target distribution when a global Nash equilibrium is reached, which requires an optimal discriminator. However, in practice, we cannot obtain such an optimal discriminator, so merely relying on discriminator loss minimization often does not imply that a target distribution has been well modeled. The network is more likely to converge around some local Nash equilibria and oscillates within a small range. The learned distribution is dynamically adjusted along with different suboptimal discriminators at different training epochs.

In image generation, inception score (IS) \cite{salimans2016improved} or Fréchet inception distance (FID) \cite{heusel2017gans} is often used as an evaluation metric to help select an optimal model checkpoint during training. Similarly, we can compute average noise power from $G_{n}$, which is expected to select a relatively optimal generator for $\mathrm{GAN}_{\, n}$ with minimum noise power errors. Moreover, we recommend performing such model selection only later in training because a GAN has not effectively converged during its initial training phase.

\begin{algorithm}
    \label{Algorithm: GN-GAN for Radio Generation}
    \caption{GN-GAN for Radio Generation}
    \SetAlgoLined
    \SetKwInOut{KIN}{Prepare}
    \KIN{$G_{\mathrm{unrolled}}(\cdot)$ with $\theta_{\scriptscriptstyle G}$ and $D(\cdot)$ with $\theta_{\scriptscriptstyle D}$}
    \vskip 0.125cm
    \textbf{Definition} $\hat{D} \coloneqq D(x) \; / \; {\Vert \nabla_x D(x)\Vert}$ \\
    \vskip 0.125cm
    \For{$i \leftarrow 1$ \KwTo $\mathrm{N}_{i}$}
    {
    \vskip 0.125cm
    \For{$j \leftarrow 1$ \KwTo $\mathrm{N}_{j}$}
    {
    \vskip 0.125cm
    sample a batch of $x$ from $p_\mathrm{data}$, e.g., $\{x_{k}\}_{k=1}^{m}$ \\
    sample a batch of $s$ from $p_s$, e.g., $\{s_{k}\}_{k=1}^{m}$ \\
    sample a batch of $\boldsymbol{z}$ from $p_{\boldsymbol{z}}$, e.g., $\{{\boldsymbol{z}}_{k}\}_{k=1}^{m}$ \\
    \vskip 0.125cm
    $\mathcal{L}_{\scriptscriptstyle D} = \frac{1}{m} \sum \limits_{k=1}^{m} { \hat{D}(G_{\mathrm{unrolled}}(\boldsymbol{z}_{k} \mid s_{k})) - \hat{D}(x_{k}) }$ \\
    \vskip 0.125cm
    $\theta_{\scriptscriptstyle D} \leftarrow \operatorname{SGD}(\nabla_{\theta_{\scriptscriptstyle D}}{\mathcal{L}_{\scriptscriptstyle D}}, \, \theta_{\scriptscriptstyle D})$ \\
    \vskip 0.125cm
    }

    \vskip 0.125cm
    sample a batch of $s$ from $p_s$, e.g., $\{s_{k}\}_{k=1}^{m}$ \\
    sample a batch of $\boldsymbol{z}$ from $p_{\boldsymbol{z}}$, e.g., $\{{\boldsymbol{z}}_{k}\}_{k=1}^{m}$ \\
    \vskip 0.125cm
    $\mathcal{L}_{\scriptscriptstyle G} = \frac{1}{m} \sum \limits_{k=1}^{m} { -\hat{D}(G_{\mathrm{unrolled}}(\boldsymbol{z}_k \mid s_k)) }$ \\
    \vskip 0.125cm
    $\theta_{\scriptscriptstyle G} \leftarrow \operatorname{SGD}(\nabla_{\theta_{\scriptscriptstyle G}}{\mathcal{L}_{\scriptscriptstyle G}}, \, \theta_{\scriptscriptstyle G})$ \\
    \vskip 0.125cm
    }
\end{algorithm}

\vspace{11pt}

\noindent\textbf{Optimization of $\mathrm{GAN}_{\, \alpha, \, n}$ and $\mathrm{GAN}_{\, h, \, n}$} Structurally, we can easily find that $\mathrm{GAN}_{\, \alpha, \, n}$ and $\mathrm{GAN}_{\, h, \, n}$ enjoy a consistent form, i.e., $u + v$, with two unknown probability distributions, $p_u$ and $p_v$, required to be learned. They are inner distributions modeled by those internal modules of an unrolled generator. The target distribution is a superimposed result of inner ones. It is not difficult to know that $p_u$ and $p_v$ correspond to $p_{as}$ and $p_{n}$ for $\mathrm{GAN}_{\, \alpha, \, n}$, respectively, or $p_{h(s)}$ and $p_{n}$ for $\mathrm{GAN}_{\, h, \, n}$.

For optimization with this form, we need to confirm an implicit premise that $p_u$ and $p_v$ must be of different distribution types. Otherwise, it can be divided into two cases: satisfy additivity or not. Such additivity refers to that two mutually independent random variables of an identical distribution type still obey a probability distribution of this type after being added together. For example, considering two mutually independent Gaussian random variables, we can see that their sum is still Gaussian, and their expectations and variances or covariance matrices also meet such an addition relationship. The additivity implies any number of feasible solutions. The learning result is bound to oscillate wildly and cannot converge effectively. Even if such additivity is not satisfied, there remains another issue when learning $p_u$ and $p_v$. The respective roles of $p_u$ and $p_v$ are equivalent, and thus they are exchangeable, which means $p_u$ might be learned as $p_v$, and accordingly, $p_v$ would be learned as $p_u$.

This issue usually does not occur in practice when $p_u$ and $p_v$ are of different distribution types because we can adopt two different neural network structures and make them sample from different latent spaces. Taking $\mathrm{GAN}_{\, h, \, n}$ as an example, where $H$ and $G_{n}$ are built with different domain knowledge. The two modules sample from a pure signal distribution $p_{s}$ and a Gaussian distribution $p_{\boldsymbol{z}}$, respectively, making it easier for $H$ to map $p_{s}$ as $p_{h(s)}$, instead of $p_{n}$. Once an initial optimization direction can be determined, $p_u$ and $p_v$ are also uniquely determined. To sum up, we believe that $p_u$ and $p_v$ should preferably belong to different distribution types. Empirically, $p_{as}$ or $p_{h(s)}$ and $p_{n}$ are generally of different distribution types.

As mentioned earlier, GANs can only converge to some local Nash equilibria in practice and oscillate within a small range. Hence, we usually need an evaluation metric to help select an optimal model checkpoint during training. However, such convergence could become even worse when there are two probability distributions required to be simultaneously learned. To this end, we present a new optimization algorithm to achieve more stable and effective convergence.

Before introducing it, we provide an auxiliary concept. In general, any probability distribution can be described with two elements: distribution type and energy level. The energy level can only be used to distinguish different distributions under an identical distribution type. For example, a Gaussian distribution has its distribution type as Gaussian, and its mean and variance or covariance matrix determine its energy level. Every possible value $\boldsymbol{x}$ of a probability distribution $p_{\boldsymbol{x}}$ corresponds to an implicit energy value $E_{\boldsymbol{x}}$, which can be defined by that energy function learned by an optimal discriminator, i.e., $E_{\boldsymbol{x}} = D^{*}(\boldsymbol{x})$. The average energy for $p_{\boldsymbol{x}}$ thus can be defined as
\begin{equation}
    E_{p_{\boldsymbol{x}}} = \sum_{\boldsymbol{x} \, \sim \, p_{\boldsymbol{x}}} D^{*}(\boldsymbol{x}) p_{\boldsymbol{x}}(\boldsymbol{x})
\end{equation}

Then, learning an unknown probability distribution has become a process of estimating its distribution type and energy level. GANs can converge to a rough result when learning $p_u$ and $p_v$ without constraints. The estimation for distribution types is relatively accurate because great morphological differences between different distribution types lead to a relatively large discriminator loss, but GANs often lack a more accurate estimation for energy levels of each inner distribution since there is no optimal discriminator available. In this case, we consider adding an explicit energy constraint for estimating energy levels better. In practice, we can take average power as an equivalent alternative for energy levels, e.g., average transmit power, average channel gain, and average noise power, which can be used to constrain $H$, $G_{\alpha}$, and $G_{n}$, respectively.

The generator and discriminator are both alternately optimized using SGD. In Algorithm \ref{Algorithm: GN-GAN for Radio Generation}, $G_{\mathrm{unrolled}}$ is optimized only one step after optimizing $D$ for many steps. Keeping this iterative fashion unchanged, we consider adding an extra optimization process after each training step of $G_{\mathrm{unrolled}}$ to optimize one of its some parts further, including $G_{\alpha}$, $G_{n}$, and $H$. Taking $\mathrm{GAN}_{\, h, \, n}$ as an example, we can apply an energy constraint for $H$ based on average transmit power. Specifically, every time after optimizing $G_{\mathrm{unrolled}}$, this energy constraint is used to optimize $H$ individually. The mean squared error (MSE) can be used as its objective function, i.e.,
\begin{equation}
    \mathcal{L} _{\mathrm{constraint}}=\left\| \mathbb{E} _{s \sim \hat{p}_s}\left\{ H(s) \right\} - c \right\| _2
\end{equation}
where $c$ denotes an expected value of average transmit power. Note that $G_{\mathrm{unrolled}}$ is frozen when $D$ is being optimized, so this energy constraint can always exist on $G_{\mathrm{unrolled}}$ whenever we optimize $D$, which is equivalent to incorporating it into the entire training process of the GAN. We summarize this optimization scheme in Algorithm \ref{Algorithm: GN-GAN for Radio Generation with Energy Constraint}. Similarly, we can apply an energy constraint based on average noise power when optimizing $\mathrm{GAN}_{\, \alpha, \, n}$, which will not be repeated here.

\begin{algorithm}
    \label{Algorithm: GN-GAN for Radio Generation with Energy Constraint}
    \caption{Energy-Constrained GAN Optimization}
    \SetAlgoLined
    \SetKwInOut{KIN}{Prepare}
    \KIN{$G_{\mathrm{unrolled}}(\cdot)$ with $\theta_{\scriptscriptstyle G}$ and $D(\cdot)$ with $\theta_{\scriptscriptstyle D}$}
    \vskip 0.125cm
    \textbf{Definition} $\hat{D} \coloneqq D(x) \; / \; {\Vert \nabla_x D(x)\Vert}$ \\
    \vskip 0.125cm
    \For{$i \leftarrow 1$ \KwTo $\mathrm{N}_{i}$}
    {
    \vskip 0.125cm
    \For{$j \leftarrow 1$ \KwTo $\mathrm{N}_{j}$}
    {
    \vskip 0.125cm
    sample a batch of $x$ from $p_\mathrm{data}$, e.g., $\{x_{k}\}_{k=1}^{m}$ \\
    sample a batch of $s$ from $p_s$, e.g., $\{s_{k}\}_{k=1}^{m}$ \\
    sample a batch of $\boldsymbol{z}$ from $p_{\boldsymbol{z}}$, e.g., $\{{\boldsymbol{z}}_{k}\}_{k=1}^{m}$ \\
    \vskip 0.125cm
    $\mathcal{L}_{\scriptscriptstyle D} = \frac{1}{m} \sum \limits_{k=1}^{m} { \hat{D}(G_{\mathrm{unrolled}}(\boldsymbol{z}_{k} \mid s_{k})) - \hat{D}(x_{k}) }$ \\
    \vskip 0.125cm
    $\theta_{\scriptscriptstyle D} \leftarrow \operatorname{SGD}(\nabla_{\theta_{\scriptscriptstyle D}}{\mathcal{L}_{\scriptscriptstyle D}}, \, \theta_{\scriptscriptstyle D})$ \\
    \vskip 0.125cm
    }

    \vskip 0.125cm
    sample a batch of $s$ from $p_s$, e.g., $\{s_{k}\}_{k=1}^{m}$ \\
    sample a batch of $\boldsymbol{z}$ from $p_{\boldsymbol{z}}$, e.g., $\{{\boldsymbol{z}}_{k}\}_{k=1}^{m}$ \\
    \vskip 0.125cm
    $\mathcal{L}_{\scriptscriptstyle G} = \frac{1}{m} \sum \limits_{k=1}^{m} { -\hat{D}(G_{\mathrm{unrolled}}(\boldsymbol{z}_k \mid s_k)) }$ \\
    \vskip 0.125cm
    $\theta_{\scriptscriptstyle G} \leftarrow \operatorname{SGD}(\nabla_{\theta_{\scriptscriptstyle G}}{\mathcal{L}_{\scriptscriptstyle G}}, \, \theta_{\scriptscriptstyle G})$ \\
    \vskip 0.125cm

    \textcolor[RGB]{192, 192, 192}{// Energy Constraint} \\
    \vskip 0.125cm
    \For{$k \leftarrow 1$ \KwTo $\mathrm{N}_{k}$}
    {
    \vskip 0.125cm
    sample a batch of $s$ from $p_s$, e.g., $\{s_{t}\}_{t=1}^{m}$ \\
    \vskip 0.125cm
    $\mathcal{L}_{\mathrm{constraint}} = \left\| \frac{1}{m} \sum \limits_{t=1}^{m} { {\lvert H(s_{t}) \rvert}^2 - c } \right\| _2$ \\
    \vskip 0.125cm
    $\theta_{\scriptscriptstyle H} \leftarrow \operatorname{SGD}(\nabla_{\theta_{\scriptscriptstyle H}}{\mathcal{L}_{\mathrm{constraint}}}, \, \theta_{\scriptscriptstyle H})$ \\
    \vskip 0.125cm
    }

    }
\end{algorithm}

\vspace{11pt}

\noindent\textbf{Optimization of $\mathrm{GAN}_{\, h, \, \alpha, \, n}$} As for $\mathrm{GAN}_{\, h, \, \alpha, \, n}$, we find it enjoys a form of $u_1 u_2 + v$, with three unknown probability distributions, $p_{u_1}$, $p_{u_2}$, and $p_v$, required to be learned, corresponding to $p_{\alpha}$, $p_{h(s)}$, and $p_{n}$, respectively. Unfortunately, this form does not have a feasible solution since $u_1$ cannot be effectively separated from $u_2$. However, benefits from our unrolled generator design, it can be solved indirectly.

Specifically, we should collect a certain amount of signal data under a stable channel condition, denoted by $\mathcal{X}_{\mathrm{stable}}$. The stable channel condition means an unchanged channel fading coefficient that gives radio signals fixed scaling and rotation. Such fixed scaling and rotation can be considered a part of transmitter characteristics, and then we can train $\mathrm{GAN}_{\, h, \, n}$ using Algorithm \ref{Algorithm: GN-GAN for Radio Generation with Energy Constraint}. As a result, we can obtain $H$ with an extra phase offset attached, expressed as $H^{*}$ for clarity. Meanwhile, $G_{n}$ has also been optimized well so that $p_{n}$ has already been learned. Only one unknown probability distribution remains, that is, $p_{\alpha}$ has yet to be learned. On this basis, we should collect more signal data under a dynamically changing channel condition, denoted by $\mathcal{X}_{\mathrm{dynamic}}$, which consists of radio samples with different channel fading coefficients. Then, we can build $\mathrm{GAN}_{\, h, \, \alpha, \, n}$ with well-trained $H^{*}$ and $G_{n}$. With $H^{*}$ and $G_{n}$ frozen, we can directly use Algorithm \ref{Algorithm: GN-GAN for Radio Generation} to train $\mathrm{GAN}_{\, h, \, \alpha, \, n}$ for learning $p_{\alpha}$. This indirect optimization scheme is summarized in Algorithm \ref{Algorithm: Indirect Optimization}.

\begin{algorithm}
    \label{Algorithm: Indirect Optimization}
    \caption{Indirect GAN Optimization}
    \SetAlgoLined
    \SetKwInOut{KIN}{Prepare}
    \KIN{Signal data with a stable channel condition, i.e., $\mathcal{X}_{\mathrm{stable}}$; Signal data with a dynamically changing channel condition, i.e., $\mathcal{X}_{\mathrm{dynamic}}$.}
    \vskip 0.125cm
    Train $\mathrm{GAN}_{\, h, \, n}$ on $\mathcal{X}_{\mathrm{stable}}$ using Algorithm \ref{Algorithm: GN-GAN for Radio Generation with Energy Constraint}, so as to build $\mathrm{GAN}_{\, h, \, \alpha, \, n}$ with well-trained $H^{*}$ and $G_{n}$. \\
    Train $\mathrm{GAN}_{\, h, \, \alpha, \, n}$ on $\mathcal{X}_{\mathrm{dynamic}}$ using Algorithm \ref{Algorithm: GN-GAN for Radio Generation}, with $H^{*}$ and $G_{n}$ frozen. \\
\end{algorithm}

\section{Experiments and Results} \label{Section: Experiments and Results}
In this section, a series of experiments are conducted to evaluate our proposed framework comprehensively.

\subsection{Data Preparation}
The signal dataset used in all our experiments is created by communication simulation, which aims to precisely control various experimental conditions (e.g., data distribution) and eliminate other possible interferences. The emitted message is random, which implies a radio signal each time consists of completely random symbols. The square-root raised cosine FIR filter with a rolloff factor of $0.35$ is used for pulse-shaping. The radio sample is of length $1024$ with $8 \times$ oversampling.

Specific transmitter characteristics can be simulated by non-linear power amplification. The nonlinearity of power amplifiers is often considered an important factor to form RF fingerprints \cite{hanna2019deep}. Aghasi \emph{et al.} \cite{aghasi2007modified} proposed an eight-coefficient model for solid-state power amplifiers (SSPA) where $A(r)$ is a function of input amplitude $r$, representing AM/AM conversion and $\varPhi(r)$, also a function of input amplitude $r$, representing AM/PM conversion, as follows.
\begin{equation}
    \left\{
    \begin{aligned}
        A(r)       & = \displaystyle \frac{\alpha_1 r^{\alpha_2} + \alpha_3 r^{\alpha_2 + 1}}{1 + \alpha_4 r^{\alpha_2 + 1}} \\
        \varPhi(r) & = \displaystyle \frac{\beta_1 r^{\beta_2} + \beta_3 r^{\beta_2 + 1}}{1 + \beta_4 r^{\beta_2 + 1}}
    \end{aligned} \right.
\end{equation}

In general, we can represent a modulated signal as
\begin{equation}
    s(t) = a(t) e^{2 \pi f_c t + \varphi(t)}
\end{equation}
where $f_c$ is carrier frequency, $a(t)$ and $\varphi(t)$ are modulated envelope and phase, respectively. The signal passes through a power amplifier based on Aghasi's model, and then we can rewrite it as
\begin{equation}
    s^{\star}(t) = A\left[a(t)\right] e^{2 \pi f_c t + \varphi(t) + \varPhi\left[a(t)\right]}
\end{equation}

Different coefficient choices could produce different power amplification characteristic curves, corresponding to different simulated devices. There is a set of real measured coefficients provided in \cite{aghasi2007modified}, as shown below.
\begin{equation}
    \left\{
    \begin{aligned}
        \boldsymbol{\alpha} & = (7.851,\,1.5388,\,-0.4511,\,6.3531)   \\
        \boldsymbol{\beta}  & = (4.6388,\,2.0949,\,-0.0325,\,10.8217)
    \end{aligned} \right.
\end{equation}
where we can apply small random offsets to yield multiple simulated devices. As listed in Table \ref{Table: Simulation with Non-Linear Power Amplification}, we have created a total of $10$ simulated devices. The other signal parameters will be stated in each experiment.

\begin{table*}[htb]
    \renewcommand\arraystretch{1.15}
    \centering
    \caption{Simulation with Non-Linear Power Amplification}
    \label{Table: Simulation with Non-Linear Power Amplification}
    \resizebox{0.7\textwidth}{!}{
        \begin{threeparttable}
            \begin{tabular}{ccc}
                \toprule
                \textbf{No.} & \textbf{Amplitude ($\boldsymbol{\alpha}$)} & \textbf{Phase ($\boldsymbol{\beta}$)}   \\
                \midrule
                \specialrule{0em}{1pt}{0pt}
                \midrule
                XMTR0        & $(10.2598,\,1.9926,\,-0.2782,\,9.5297)$    & $(6.0838,\,1.3190,\,-0.0375,\,16.2325)$ \\
                XMTR1        & $(10.7344,\,2.0668,\,-0.5015,\,9.5297)$    & $(6.3304,\,1.3058,\,-0.0348,\,16.2325)$ \\
                XMTR2        & $(11.6849,\,2.0193,\,-0.6689,\,9.5297)$    & $(6.7758,\,2.0689,\,-0.0280,\,16.2325)$ \\
                XMTR3        & $(10.2963,\,1.7932,\,-0.2929,\,9.5297)$    & $(6.1256,\,1.4660,\,-0.0297,\,16.2325)$ \\
                XMTR4        & $(11.3625,\,2.0100,\,-0.4304,\,9.5297)$    & $(6.6729,\,2.2441,\,-0.0168,\,16.2325)$ \\
                XMTR5        & $(11.4996,\,2.0766,\,-0.5835,\,9.5297)$    & $(6.7440,\,2.9490,\,-0.0454,\,16.2325)$ \\
                XMTR6        & $(10.5223,\,1.7999,\,-0.5658,\,9.5297)$    & $(6.4241,\,1.4531,\,-0.0425,\,16.2325)$ \\
                XMTR7        & $(10.4870,\,1.8997,\,-0.4515,\,9.5297)$    & $(6.4135,\,1.4193,\,-0.0305,\,16.2325)$ \\
                XMTR8        & $(11.3525,\,2.2360,\,-0.2442,\,9.5297)$    & $(6.9513,\,2.1135,\,-0.0366,\,16.2325)$ \\
                XMTR9        & $(10.0237,\,1.9307,\,-0.4582,\,9.5297)$    & $(6.0633,\,2.4454,\,-0.0363,\,16.2325)$ \\
                \bottomrule
            \end{tabular}
        \end{threeparttable}}
\end{table*}

\subsection{Benchmark of Radio Recognition} \label{Benchmark of Radio Recognition}
To begin with, we perform a benchmark on two typical radio recognition tasks, i.e., modulation recognition and RF fingerprinting, which aims to study how channel effects impact radio recognition. In modulation recognition, with $\mathrm{Signal}_{\, n}$ and $\mathrm{Signal}_{\, \alpha, \, n}$ for communication simulation, we consider $6$ commonly used digital modulation types: BPSK, QPSK, 8PSK, 16QAM, 32QAM, and 64QAM. In RF fingerprinting, with $\mathrm{Signal}_{\, h, \, n}$ and $\mathrm{Signal}_{\, h, \, \alpha, \, n}$ for communication simulation, we consider those $10$ simulated devices given in Table \ref{Table: Simulation with Non-Linear Power Amplification} and adopt QPSK modulation.

The average transmit power is set to $1.0$, and consider a Rayleigh fading channel with white Gaussian noise added, or only consider additive white Gaussian noise (AWGN). The average channel gain for Rayleigh fading is set to $1.0$. The (average) signal-to-noise ratio (SNR) is set to $18$ dB. There are a total of $9000$ radio samples provided for each class. The signal data is then divided into a training set, a validation set, and an evaluation set in a proportion of $5:2:2$.

\begin{figure}[htb]
    \centering
    \includegraphics[scale=0.8]{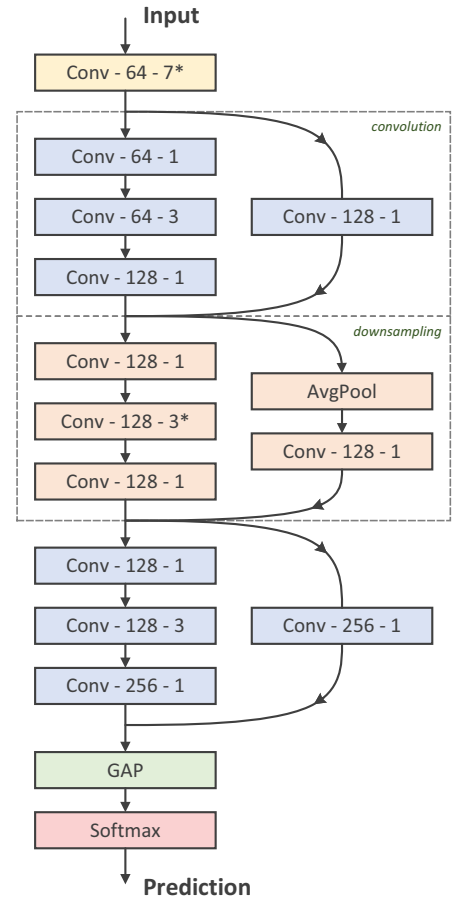}
    \caption{The structure of ResNet we used for radio recognition. The layer label ``Conv-$k$-$s$'' means this is a convolutional layer that has $k$ kernels of size $s$. All convolutions of size $3$ are separable. The asterisk (*) indicates that this layer adopts a stride of size $2$. Batch normalization (BN) \cite{ioffe2015batch} has been performed after each convolution, followed by non-linear activation using rectified linear units (ReLU) \cite{nair2010rectified}.}
    \label{Figure: Network}
\end{figure}

The classifier we used is a deep residual network (ResNet) \cite{he2016deep}, as illustrated in Fig. \ref{Figure: Network}. The network starts from a convolutional layer, followed by a series of alternately stacked convolution and downsampling blocks, and ends with a classification layer. The initial convolutional layer is used as an input stem, and it has $64$ convolution kernels of size $7$, with a stride of $2$. The convolution and downsampling blocks are both implemented as residual blocks, and their designs mainly refer to ResNet-B and ResNet-D mentioned in \cite{he2019bag}. The last classification layer performs global average pooling (GAP) and then yields a prediction using softmax. The number of residual blocks can be varied to obtain different ResNet models. Here we adopt two contiguous convolution blocks with a downsampling block between them as transition. The model is built with TensorFlow \cite{tensorflow2015} and then trained on a single NVIDIA RTX 2080S GPU utilizing an Adam \cite{kingma2014adam} solver for $180$ epochs with a mini-batch size of $128$. The initial learning rate is set to $0.001$. The recognition accuracy is used as a performance metric and calculated based on $20$ trials.

\begin{figure}[htb]
    \centering
    \includegraphics[scale=0.625]{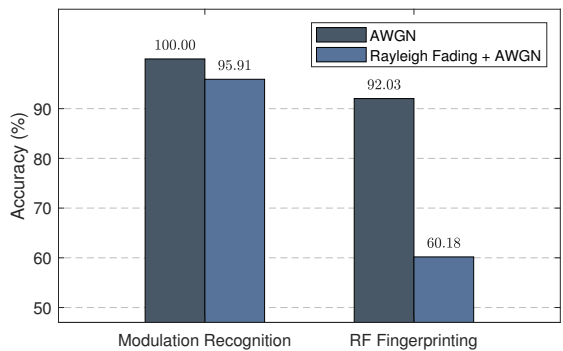}
    \caption{Benchmark of radio recognition.}
    \label{Figure: Benchmark of Radio Recognition}
\end{figure}

The experimental results are presented in Fig. \ref{Figure: Benchmark of Radio Recognition}, where we compare radio recognition performance when experiencing with and without channel fading. It can be seen that modulation recognition is relatively less affected by channel fading. The reason is that modulation recognition largely depends on constellation topology differences to make a decision. It is knowledgeable that radio signals of different modulation types have different constellation topologies. As for a single radio sample, we often assume that channel characteristics are invariant during such a short time, and channel fading only causes random rotation and scaling for its constellation topology but hardly changes its topological morphology itself. The only impact is that SNR may decrease to some extent because channel fading only reduces signal power but does not change noise power. The overall recognition accuracy thus decreases slightly.

\begin{figure*}[htb]
    \centering
    \subfigure[$58.27\%$]
    {
        \centering
        \includegraphics[width=0.3\textwidth]{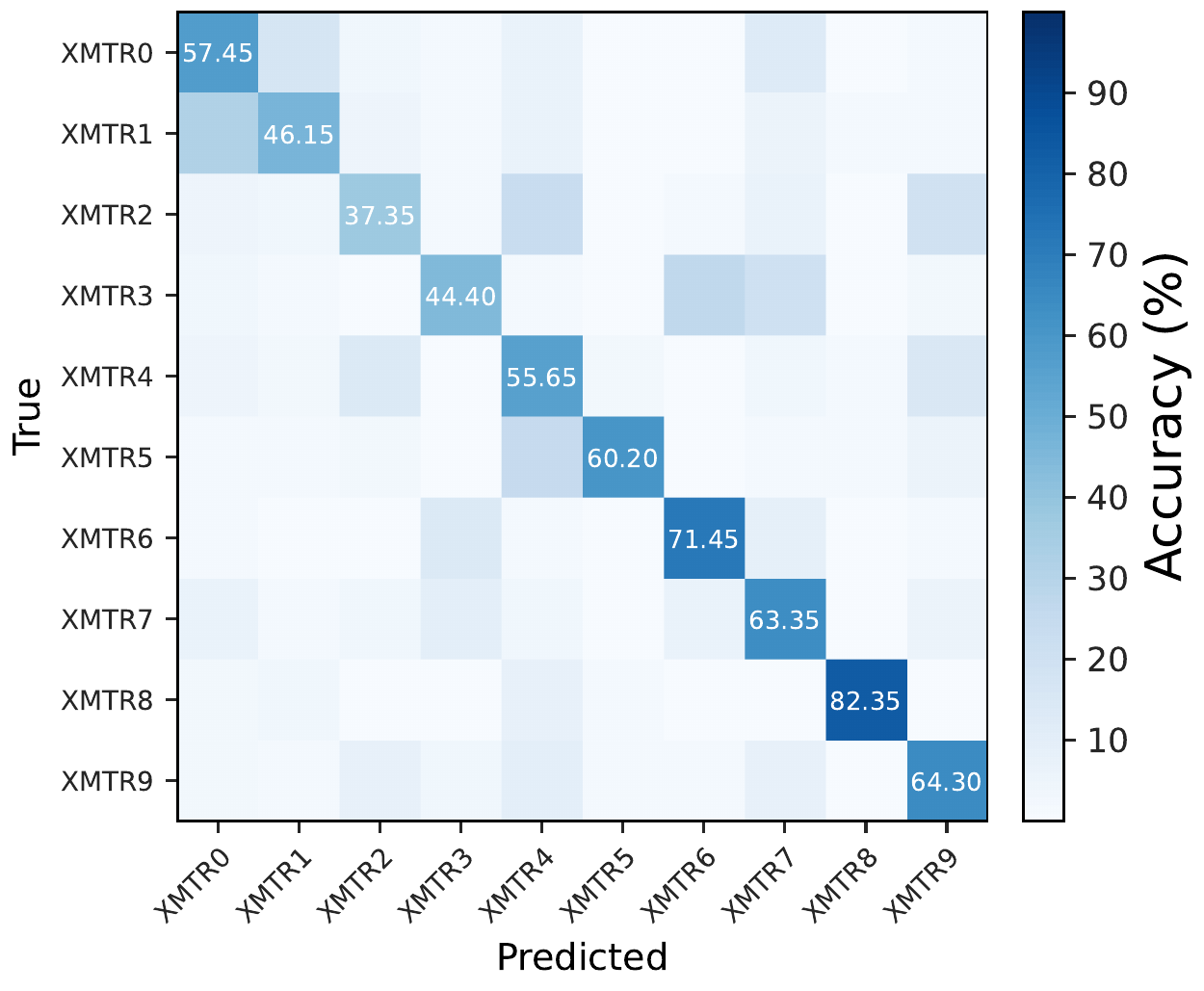}
    }
    \hspace{0.75em}
    \subfigure[$61.80\%$]
    {
        \centering
        \includegraphics[width=0.3\textwidth]{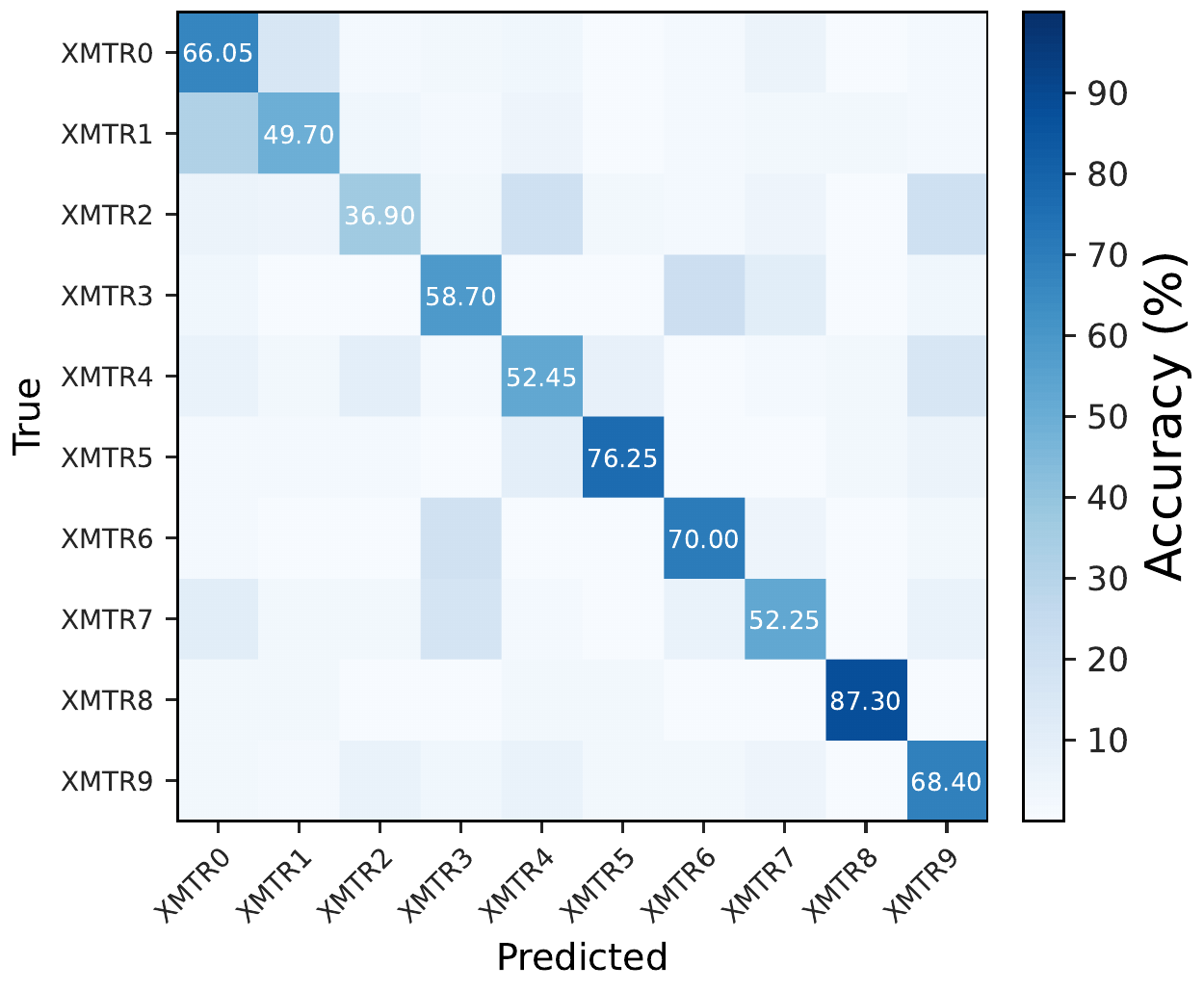}
    }
    \hspace{0.75em}
    \subfigure[$60.10\%$]
    {
        \centering
        \includegraphics[width=0.3\textwidth]{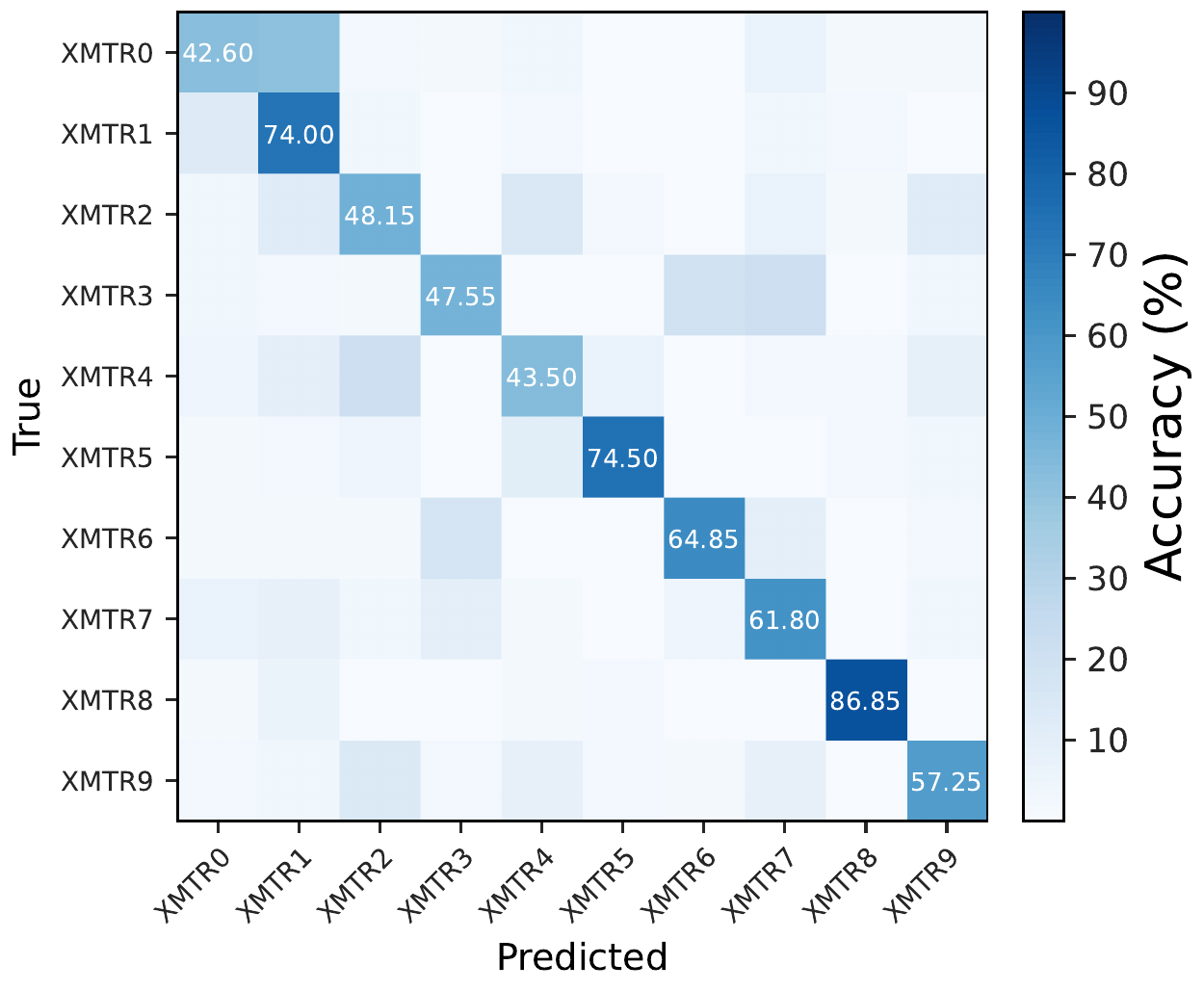}
    }
    \caption{Confusion matrices obtained on an identical dataset ($\mathrm{Signal}_{\, h, \, \alpha, \, n}$) with three different trials.}
    \label{Figure: Confusion Matrix}
\end{figure*}

However, we can see that channel fading severely impacts RF fingerprinting. The overall recognition accuracy has deteriorated from $92.03\%$ to only about $60\%$ when experiencing channel fading. More importantly, such classification results may no longer be reliable. As shown in Fig. \ref{Figure: Confusion Matrix}, we have presented three confusion matrices obtained on an identical dataset with different trials. The average recognition accuracy is no longer stable for each device, which means our classifier would yield inconsistent results if we randomly select some batches of radio samples for evaluation, even if they all follow an identical sample distribution. Furthermore, we can see that many devices (e.g., XMTR0) achieve an average recognition accuracy of less than $50\%$, which usually makes no sense in practice. This result also partly reflects that $\mathrm{GAN}_{\, h, \, \alpha, \, n}$ cannot be solved directly.

\subsection{Visualization of Learning Results}
The inner trainable modules in our proposed GAN framework, including $G_{\alpha}$, $G_{n}$, and $H$, all essentially learn a complex-valued probability distribution. The target distribution to be modeled, i.e., an underlying sampling distribution of radio signals, is also complex-valued. More importantly, they are all one-dimensional. It is not difficult to intuitively visualize learning results via a density histogram or density spectrum using non-parametric density estimation \cite{scott2015multivariate}.

The related signal parameters for communication simulation are basically consistent with what we previously used for classification benchmark. In particular, only one device is considered when simulating specific transmitter characteristics. The real measured coefficients is used. The modulation type is configured as QPSK. The other parameters (e.g., average transmit power) remain unchanged unless otherwise specified. There are sufficient radio samples provided, from which we have split $1024 \times 256$ sampling points for training. The related GANs, i.e., $\mathrm{GAN}_{\, n}$, $\mathrm{GAN}_{\, \alpha, \, n}$, $\mathrm{GAN}_{\, h, \, n}$, and $\mathrm{GAN}_{\, h, \, \alpha, \, n}$, are also built with TensorFlow and then trained using an Adam solver for $10000$ epochs with a mini-batch size of $1024$ (sampling points). The initial learning rate is set to $0.0003$.

\begin{figure*}[htb]
    \centering
    \subfigure[18 dB, $\mathrm{P}_{n} = 0.01585$, $\hat{\mathrm{P}}_{n} = 0.01587$]
    {
        \centering
        \includegraphics[width=0.275\textwidth]{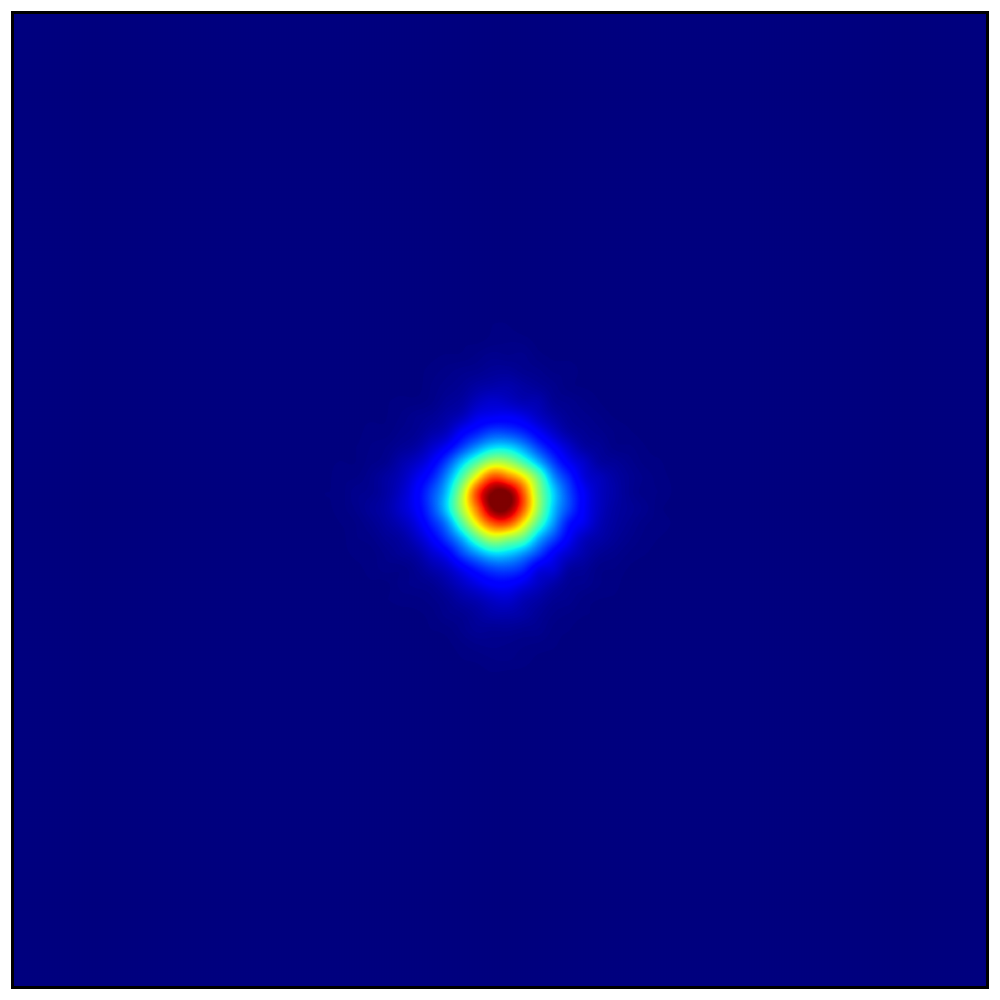}
    }
    \hspace{1.8em}
    \subfigure[12 dB, $\mathrm{P}_{n} = 0.06310$, $\hat{\mathrm{P}}_{n} = 0.06332$]
    {
        \centering
        \includegraphics[width=0.275\textwidth]{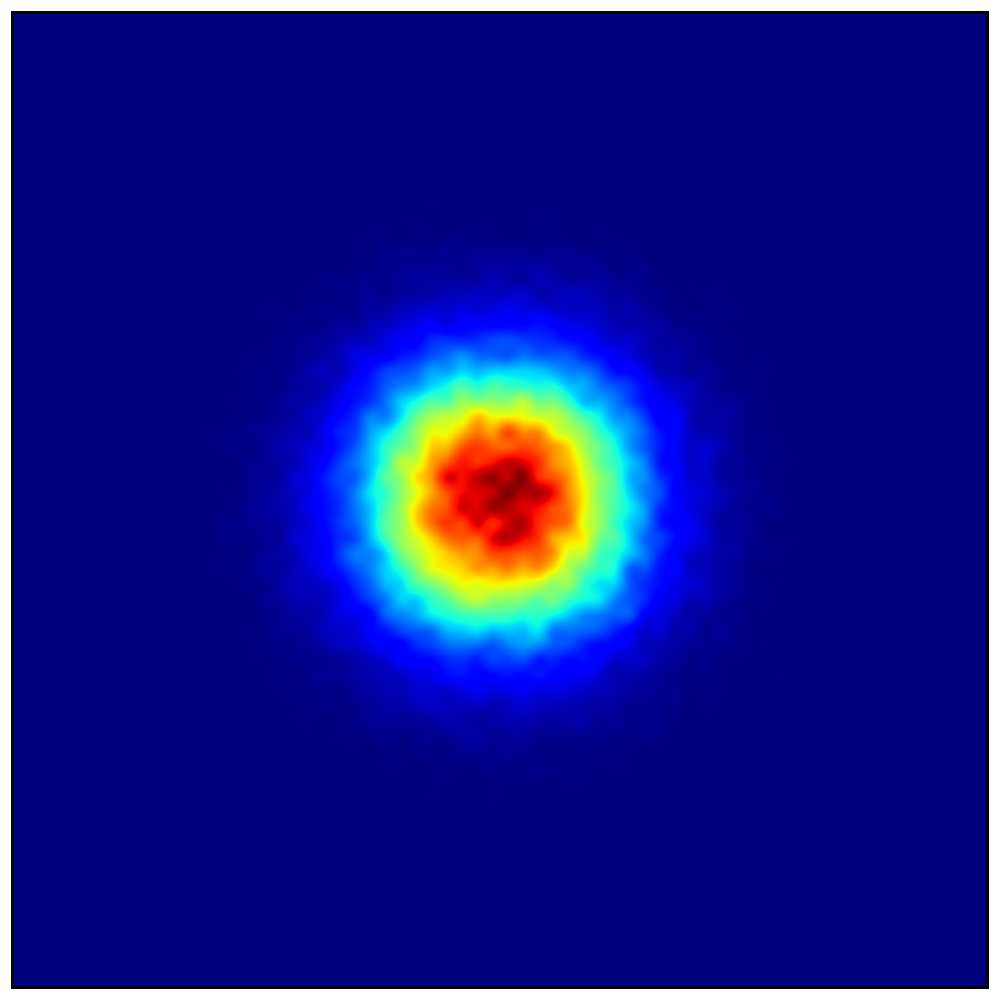}
    }
    \hspace{1.8em}
    \subfigure[6 dB, $\mathrm{P}_{n} = 0.25119$, $\hat{\mathrm{P}}_{n} = 0.26317$]
    {
        \centering
        \includegraphics[width=0.275\textwidth]{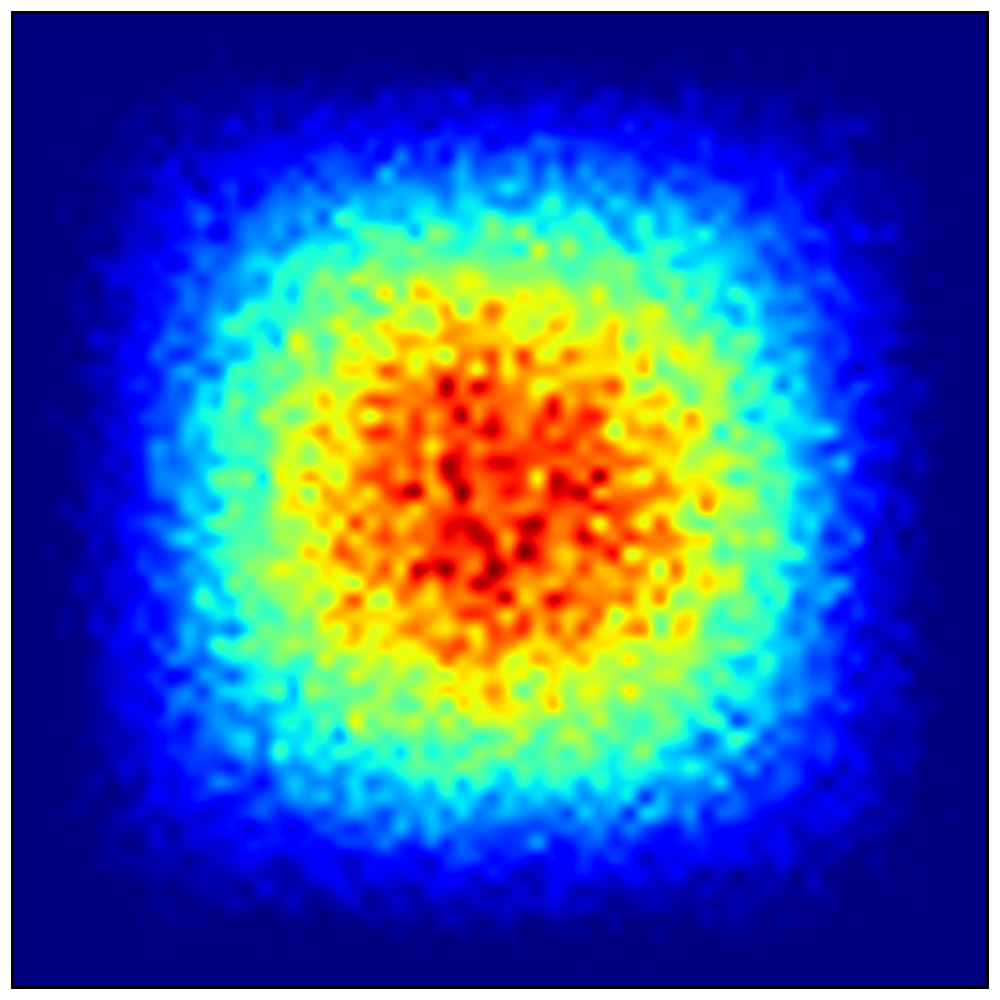}
    }
    \caption{The learning results of $\mathrm{GAN}_{\, n}$. The distribution of noise ($p_{n}$) is plotted with a density spectrum. The selected region for non-parametric density estimation is $[-2, \, 2)^2$ with $64$ bins divided on each side, so we can obtain a density spectrum of size $64 \times 64$. Note that $\mathrm{P}_{n}$ is a theoretical value for average noise power, while $\hat{\mathrm{P}}_{n}$ is estimated from generated data.}
    \label{Figure: Noise}
\end{figure*}

The learning results of $\mathrm{GAN}_{\, n}$ optimized with Algorithm \ref{Algorithm: GN-GAN for Radio Generation} are shown in Fig. \ref{Figure: Noise}, where we consider three SNR conditions: 18 dB (high), 12 dB (medium), and 6 dB (low). The noise distribution learned by $G_{n}$, we can see that, is approximately Gaussian, and its power is also estimated accurately. In addition, we observe that this estimation error tends to increase with a decrease in SNRs.

\begin{figure}[htb]
    \centering
    \includegraphics[scale=0.625]{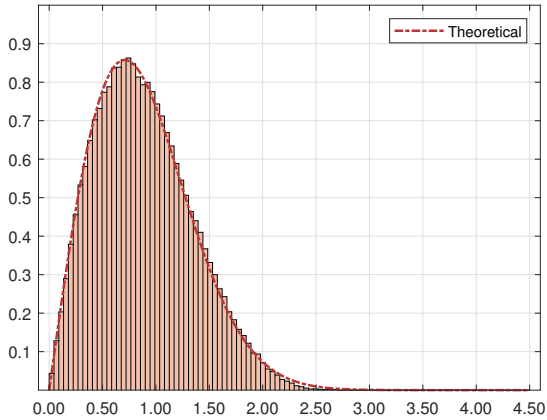}
    \caption{The learning result of $\mathrm{GAN}_{\, \alpha, \, n}$. The distribution of channel fading ($p_{a}$) is plotted with a density histogram. The average channel gain estimated from generated data is $1.00273$.}
    \label{Figure: Channel Fading}
\end{figure}

The learning result of $\mathrm{GAN}_{\, \alpha, \, n}$ optimized using Algorithm \ref{Algorithm: GN-GAN for Radio Generation with Energy Constraint} with $G_{n}$ constrained on average noise power is shown in Fig. \ref{Figure: Channel Fading}, where we only plot its learning result for channel fading since that for noise is very similar to $\mathrm{GAN}_{\, n}$. Hence, we no longer repeatedly and redundantly exhibit it. The channel fading learned by $G_{\alpha}$, we can see that, indeed follows a Rayleigh distribution with an enough accurate estimation for average channel gain.

\begin{figure}[htb]
    \centering
    \subfigure[In-Phase]
    {
        \centering
        \includegraphics[scale=0.625]{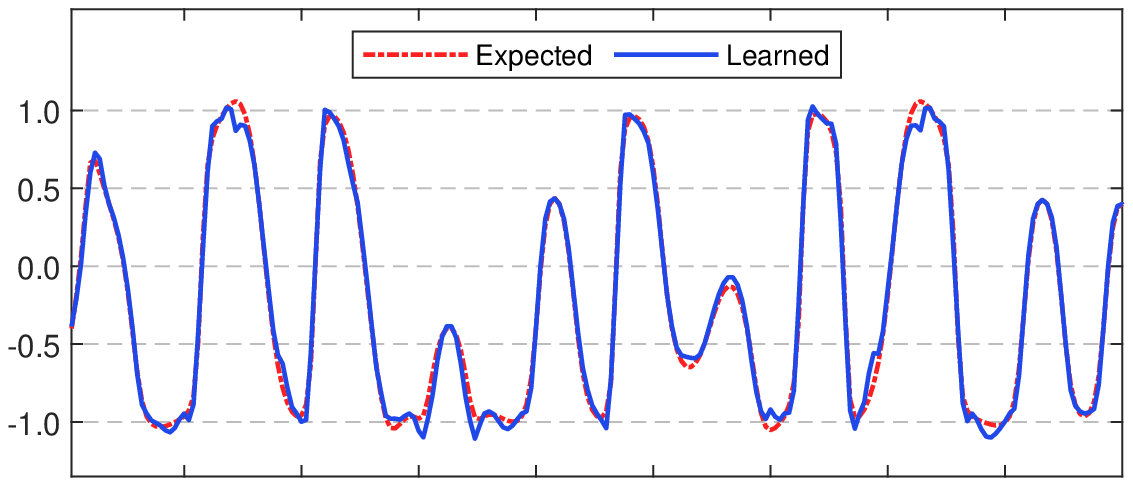}
    }
    \\
    \subfigure[Quadrature]
    {
        \centering
        \includegraphics[scale=0.625]{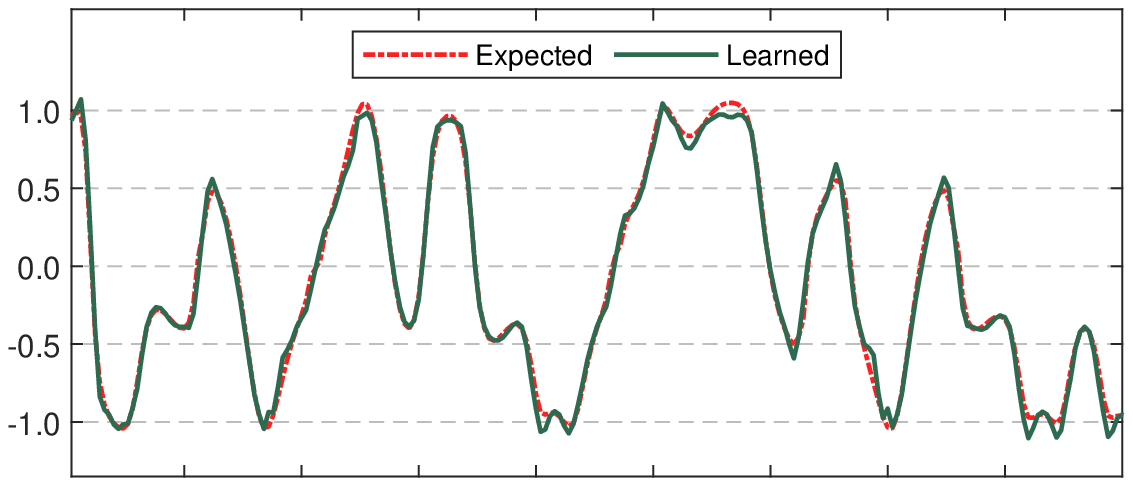}
    }
    \caption{The learning result of $\mathrm{GAN}_{\, h, \, n}$. The output of $H$ is plotted.}
    \label{Figure: RF Fingerprints}
\end{figure}

The learning result of $\mathrm{GAN}_{\, h, \, n}$ optimized using Algorithm \ref{Algorithm: GN-GAN for Radio Generation with Energy Constraint} with $H$ constrained on average transmit power is shown in Fig. \ref{Figure: RF Fingerprints}, where a pure signal waveform after experiencing learned transmitter characteristics is compared with its expected one. It can be seen that they are highly consistent, which demonstrates that specific transmitter characteristics have been correctly captured.

\subsection{Comparison with Other GANs}
The proposed framework is also compared with conventional GANs. As stated earlier, in modulation recognition, Patel \emph{et al.} \cite{patel2020data} and Lee \emph{et al.} \cite{lee2021uniqgan} exploited a CGAN and a variant of ACGAN for signal data augmentation, respectively. Generally, CGANs and ACGANs can be regarded as representatives of conventional GANs. In practice, we select a better conventional GAN design for comparison, CGANs with a projection discriminator \cite{miyato2018cgans}. The dataset is still created by communication simulation, where $\mathrm{Signal}_{\, n}$, $\mathrm{Signal}_{\, \alpha, \, n}$, and $\mathrm{Signal}_{\, h, \, n}$ are considered, whereas $\mathrm{Signal}_{\, h, \, \alpha, \, n}$ is not, because channel fading can seriously affect RF fingerprinting, as proved earlier. The related parameters for communication simulation are consistent with our previous classification benchmark.

\vspace{11pt}

\begin{figure}[htb]
    \centering
    \includegraphics[scale=0.7]{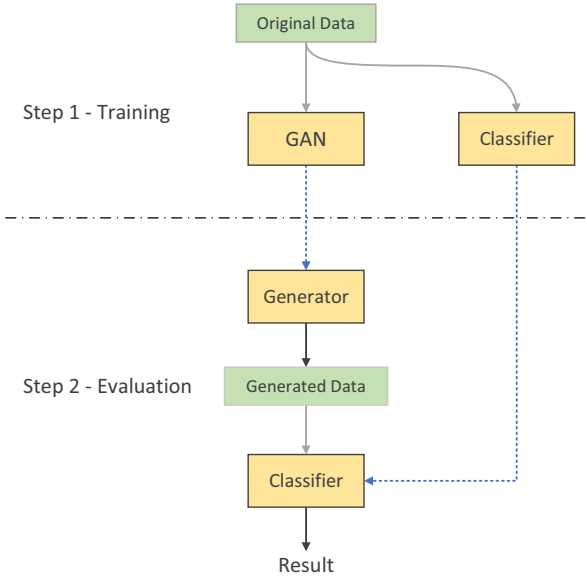}
    \caption{Evaluation of modeling effects for GANs.}
    \label{Figure: Flow 1}
\end{figure}

\begin{figure}[htb]
    \centering
    \subfigure[Conventional GANs, $88.53\%$]
    {
        \centering
        \includegraphics[width=0.225\textwidth]{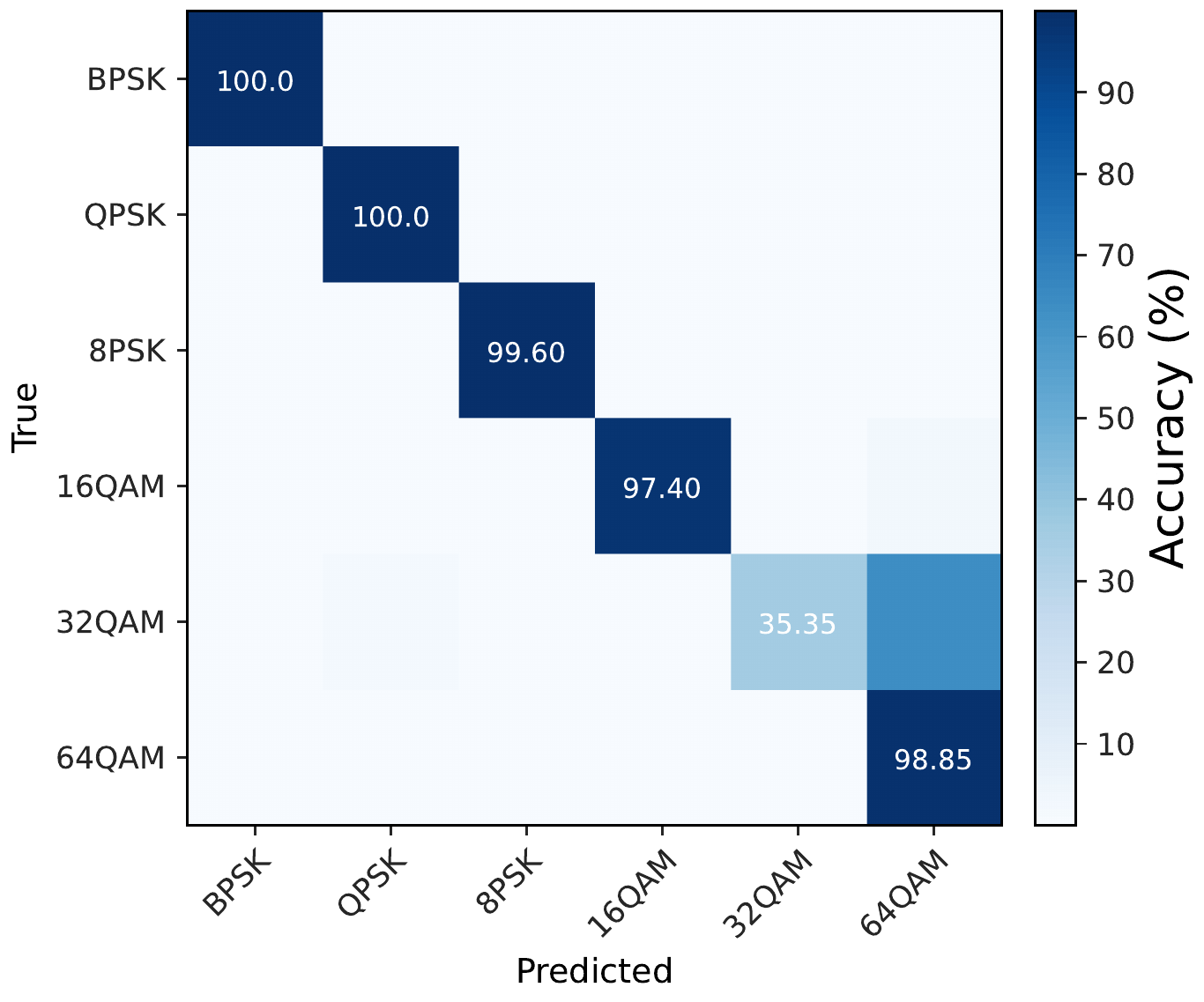}
    }
    \hspace{-0.15em}
    \subfigure[Proposal, $100.0\%$]
    {
        \centering
        \includegraphics[width=0.225\textwidth]{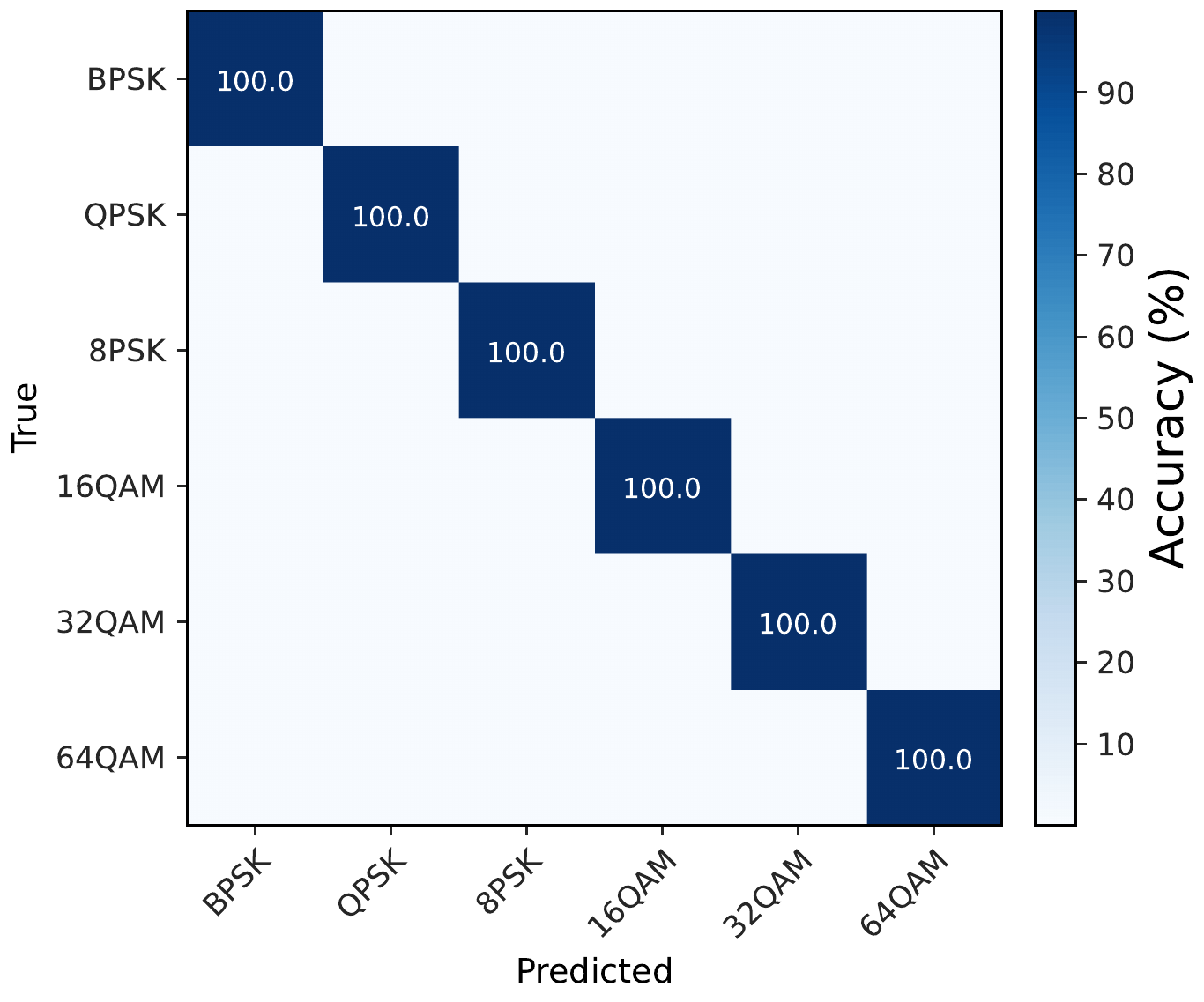}
    }
    \\
    \subfigure[Conventional GANs, $77.85\%$]
    {
        \centering
        \includegraphics[width=0.225\textwidth]{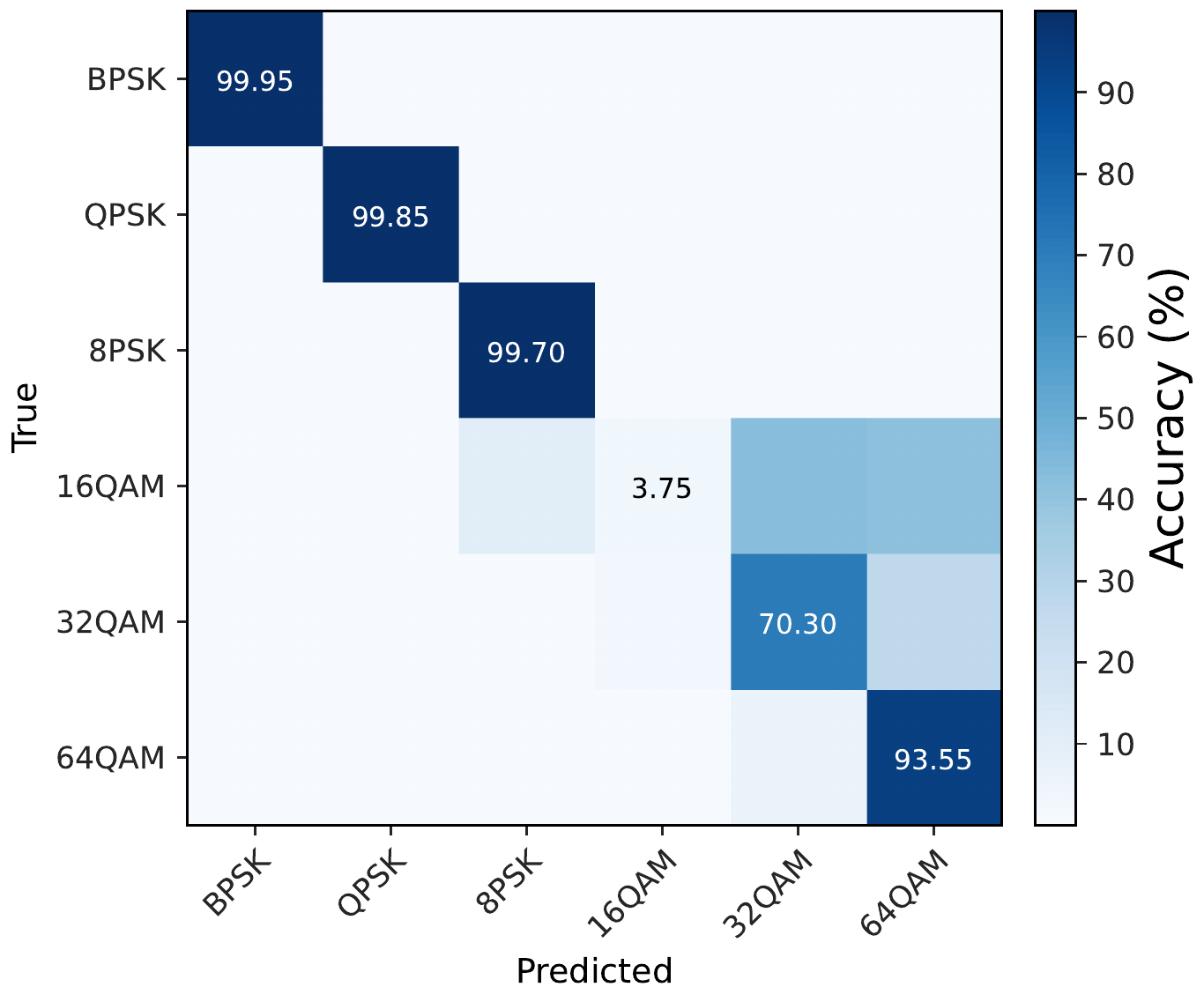}
    }
    \hspace{-0.15em}
    \subfigure[Proposal, $95.68\%$]
    {
        \centering
        \includegraphics[width=0.225\textwidth]{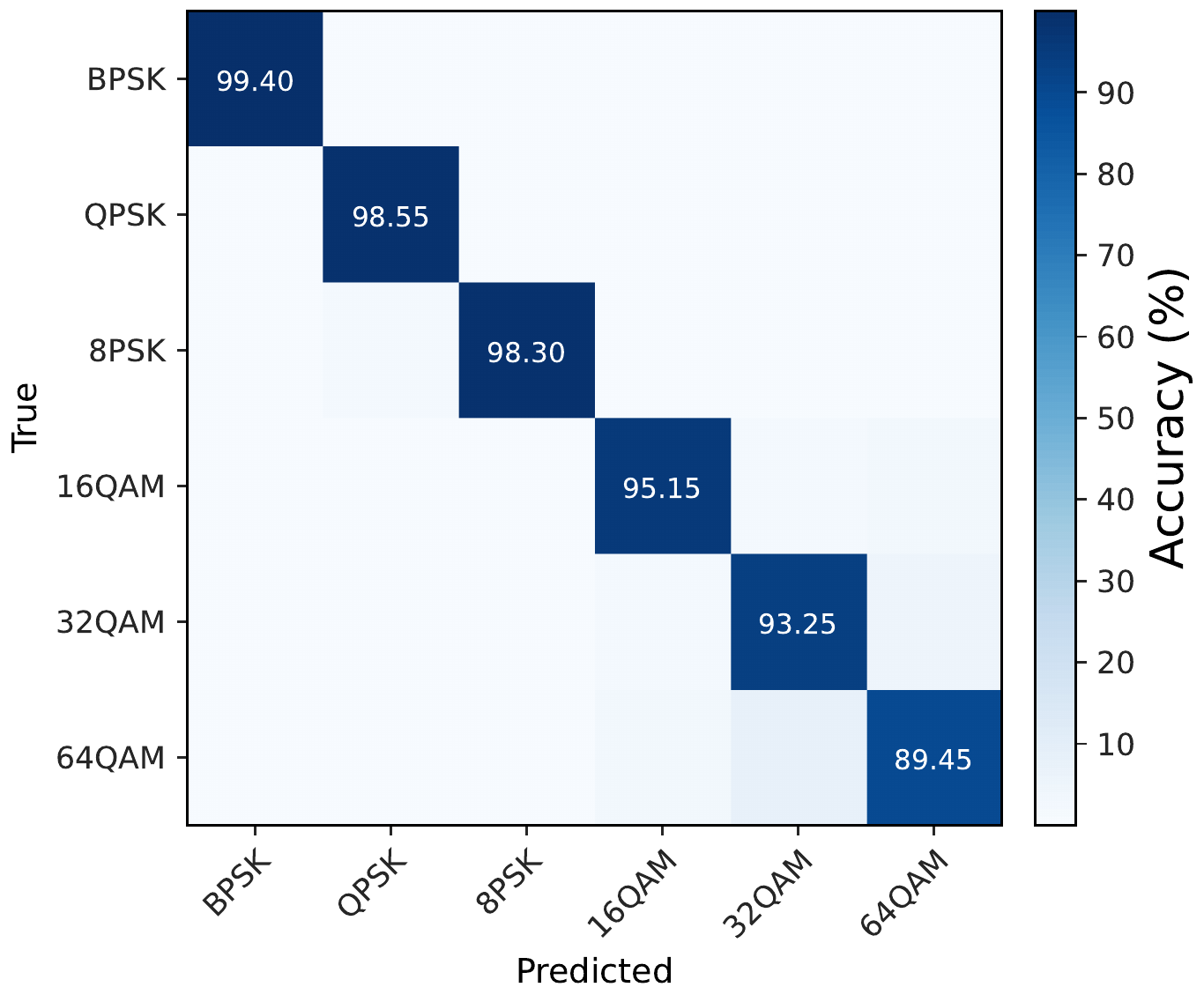}
    }
    \\
    \subfigure[Conventional GANs, $35.86\%$]
    {
        \centering
        \includegraphics[width=0.225\textwidth]{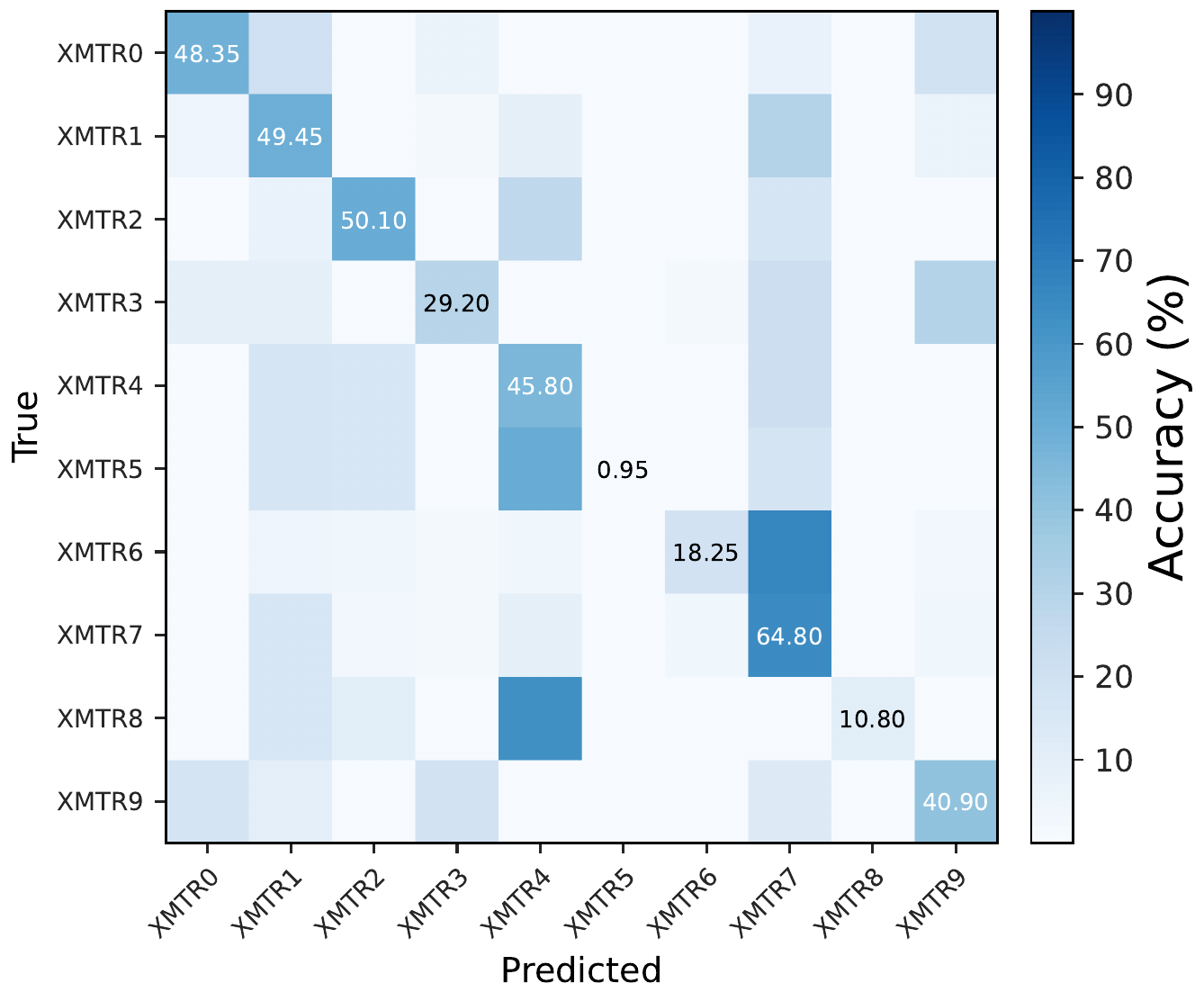}
    }
    \hspace{-0.15em}
    \subfigure[Proposal, $91.86\%$]
    {
        \centering
        \includegraphics[width=0.225\textwidth]{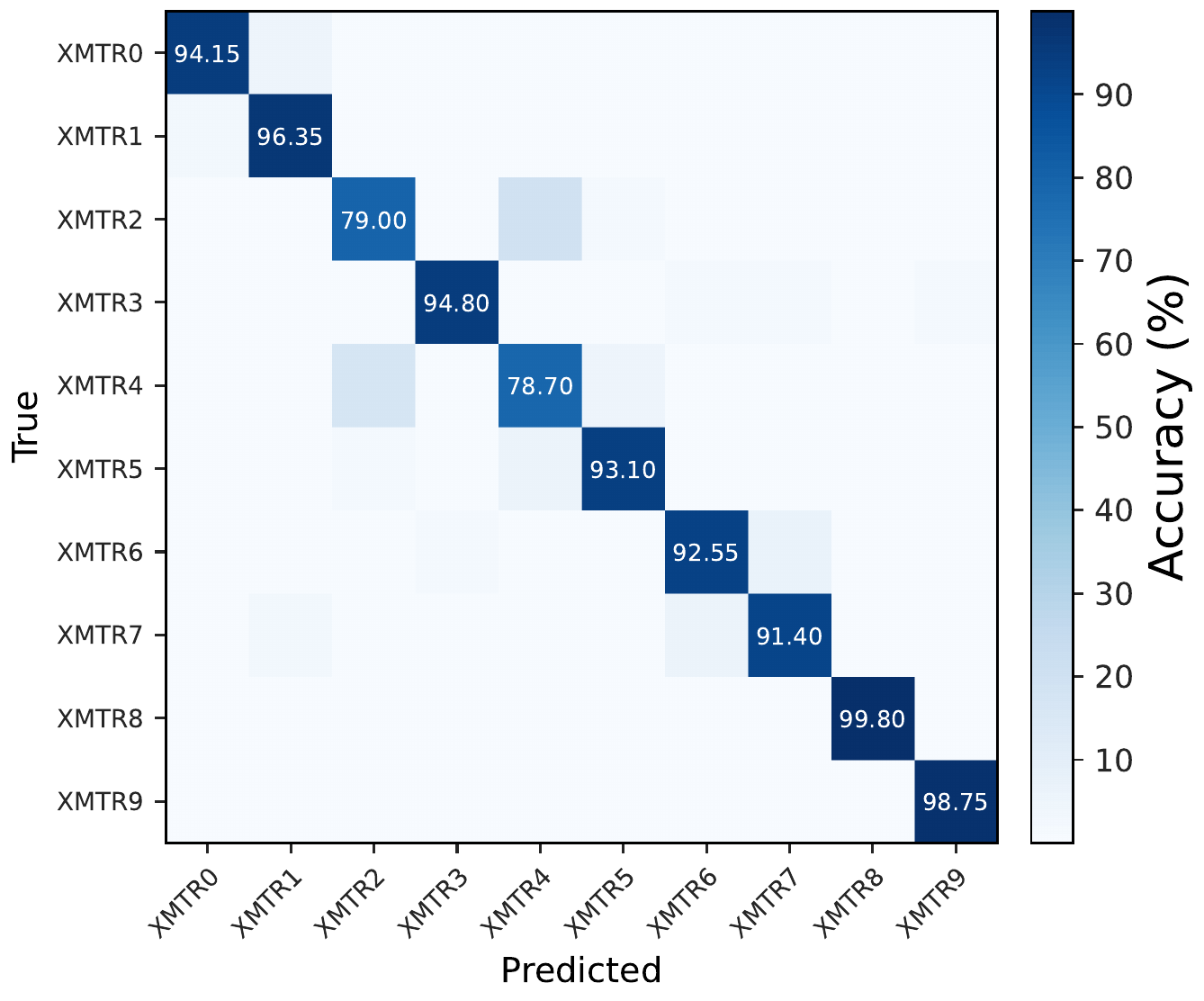}
    }
    \caption{Comparisons between conventional GANs and our proposal via classifying on generated data. $\mathrm{Signal}_{\, n}$ is used to create a signal dataset for modulation recognition in (a) and (b). Similarly, $\mathrm{Signal}_{\, \alpha, \, n}$ is used in (c) and (d), and $\mathrm{Signal}_{\, h, \, n}$ is used in (e) and (f).}
    \label{Figure: Comparsion of Classification Effects}
\end{figure}

\noindent\textbf{Comparison of Modeling Effects} At first, we will compare different GANs in terms of their modeling effects for signal data. The specific flow is illustrated in Fig. \ref{Figure: Flow 1}, where a GAN and a classifier (i.e., ResNet) are individually trained on an identical dataset. The trained GAN, more precisely, its generator, is employed to synthesize a certain number of radio samples ($2000$ samples per class). The trained classifier classifies these generated samples to obtain a confusion matrix. If a GAN can successfully model a radio sample distribution, in theory, we can obtain a classification result on its generated samples that is very close to classifying real samples. The confusion matrix thus can largely reflect how effective a GAN has learned for each class of signal data, as given in Fig. \ref{Figure: Comparsion of Classification Effects}.

For modulation recognition, we can see that signal data of higher-order modulations have been poorly modeled with conventional GANs. The generated samples thus have been severely misclassified. The main reason is that higher-order modulations have small symbol spacings, which implies a relatively complex constellation topology. The underlying sampling distribution corresponding to such constellation topologies also becomes complex and easily confused when affected by noise. Moreover, we can see that it becomes even worse when affected by channel fading. In contrast, ours achieves a result nearly close to classifying on real samples. The estimated pure signal distribution and our unrolled generator design play an important role.

This evaluation method has a relatively high error tolerance for modulation recognition, which involves a matter of modeling precision. Specifically, even though conventional GANs cannot model for signal data of a certain modulation type very accurately, their generated samples of this modulation type may still be correctly classified, as long as they are roughly approximated, since many modulation types (e.g., BPSK, QPSK, 8PSK) have very significant differences. In contrast, we argue that using RF fingerprinting for evaluation can better reflect modeling effects since all our simulated devices exhibit very small RF fingerprint differences, meaning their waveforms are very similar. If a GAN cannot model such differences accurately, there must be serious misclassification with its generated samples. As expected, ours is still far superior to conventional GANs when considering RF fingerprinting for evaluation. Nevertheless, many real-world devices may produce more prominent RF fingerprints because of different models, electronics, manufacturers, and signal parameters, as well as other possible factors, making conventional GANs may not perform too badly.

\vspace{11pt}

\begin{figure}[htb]
    \centering
    \includegraphics[scale=0.7]{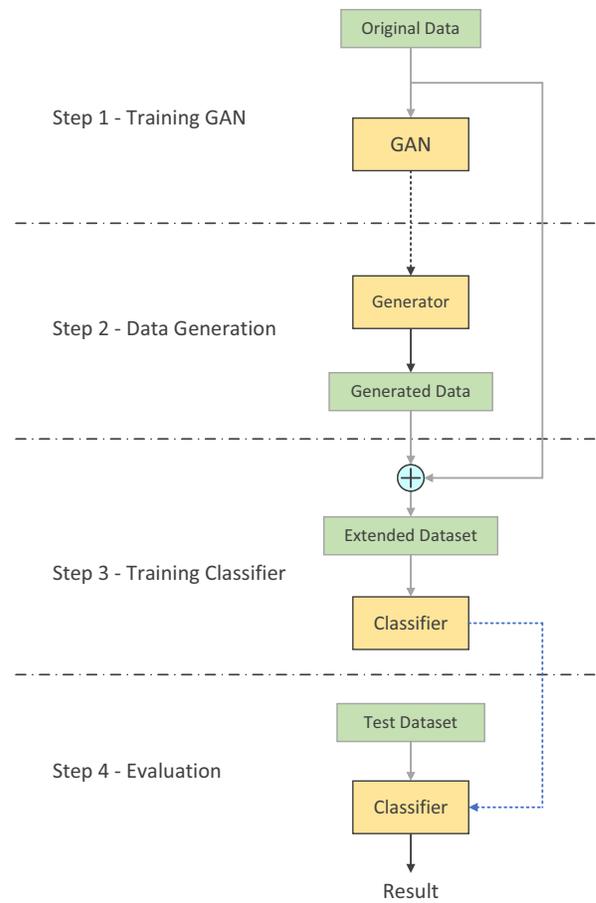}
    \caption{Using GANs for data augmentation.}
    \label{Figure: Flow 2}
\end{figure}

\noindent\textbf{Comparison of Data Augmentation} After that, we make a further comparison based on data augmentation, where only a very limited number of labeled samples are available for training. The specific flow is illustrated in Fig. \ref{Figure: Flow 2}. The trained GAN, more precisely, its generator, is employed to synthesize more labeled samples, mixed with these original ones, to train a classifier. The trained classifier is evaluated using real data. The results are listed in Table \ref{Table: Comparsion with Other GANs for Data Augmentation}. The results with rotation and flipping for signal data augmentation \cite{huang2019data} are also presented.

It can be seen that conventional GANs do not achieve any effective augmentation effects. There are two main reasons. First, as proved above, conventional GANs cannot accurately model a radio sample distribution, even with an ideal number of samples provided for training, especially with complex cases. The other is that training such a GAN for data augmentation itself requires a certain amount of labeled data, whereas a small dataset may not allow conventional GANs to be optimized well. In this case, many invalid samples are synthesized by conventional GANs, and introducing so many invalid samples could harm classifier training, thus further deteriorating performance. In contrast, ours achieves a result nearly close to fully supervised learning, though it depends on a good estimation of related signal parameters to build a pure signal distribution. The strategy of learning based on sampling points undoubtedly plays an important role, significantly reducing demands for training data so that our GANs can still be trained well even with a limited number of samples provided.

\begin{table*}[htb]
    \renewcommand\arraystretch{1.12}
    \centering
    \caption{Comparsion with Other Methods for Data Augmentation}
    \label{Table: Comparsion with Other GANs for Data Augmentation}
    \resizebox{0.7\textwidth}{!}{
        \begin{threeparttable}
            \begin{tabular}{cccccccccc}
                \toprule
                \multirow{3}{*}[-1.35ex]{\textbf{Method}}
                 & \multicolumn{6}{c}{\textbf{Modulation Recognition}}
                 & \multicolumn{3}{c}{\textbf{RF Fingerprinting}}                       \\
                \cmidrule(lr){2-7} \cmidrule(lr){8-10}
                 & \multicolumn{3}{c}{$\mathrm{Signal}_{\, n} \mid 100.0\%$}
                 & \multicolumn{3}{c}{$\mathrm{Signal}_{\, \alpha, \, n} \mid 95.91\%$}
                 & \multicolumn{3}{c}{$\mathrm{Signal}_{\, h, \, n} \mid 92.03\%$}
                \\
                \cmidrule(lr){2-4} \cmidrule(lr){5-7} \cmidrule(lr){8-10}
                 & $\boldsymbol{10}$
                 & $\boldsymbol{20}$
                 & $\boldsymbol{100}$
                 & $\boldsymbol{10}$
                 & $\boldsymbol{20}$
                 & $\boldsymbol{100}$
                 & $\boldsymbol{10}$
                 & $\boldsymbol{20}$
                 & $\boldsymbol{100}$                                                   \\
                \midrule
                Baseline (w/o)
                 & $82.10$
                 & $92.35$
                 & $99.53$
                 & $57.12$
                 & $65.13$
                 & $83.79$
                 & $26.53$
                 & $40.12$
                 & $73.10$                                                              \\
                \midrule
                Rotation
                 & $95.31$
                 & $98.65$
                 & $100.0$
                 & $68.10$
                 & $80.56$
                 & $93.25$
                 & $52.73$
                 & $67.08$
                 & $84.95$                                                              \\
                Flipping
                 & $94.52$
                 & $98.10$
                 & $100.0$
                 & $65.78$
                 & $76.25$
                 & $92.72$
                 & $32.22$
                 & $46.38$
                 & $74.67$                                                              \\
                Rotation + Flipping
                 & $97.22$
                 & $99.30$
                 & $100.0$
                 & $78.53$
                 & $86.38$
                 & $93.57$
                 & $48.76$
                 & $63.55$
                 & $83.22$                                                              \\
                \midrule
                Conventional GANs
                 & $71.26$
                 & $83.67$
                 & $94.68$
                 & $51.36$
                 & $58.90$
                 & $68.37$
                 & $21.62$
                 & $32.58$
                 & $63.63$                                                              \\
                Proposal
                 & $100.0$
                 & $100.0$
                 & $100.0$
                 & $93.86$
                 & $94.33$
                 & $95.06$
                 & $91.32$
                 & $91.65$
                 & $91.80$                                                              \\
                \bottomrule
            \end{tabular}
            \begin{tablenotes}
                \footnotesize
                \item The table headers like ``$\boldsymbol{100}$'' means there are $100$ labeled samples per class provided for training.
            \end{tablenotes}
        \end{threeparttable}}
\end{table*}

\section{Conclusion} \label{Section: Conclusion}
This paper has developed a novel GAN framework for radio generation, which can effectively learn transmitter characteristics and various channel effects, thus accurately modeling an underlying sampling distribution of radio signals. In future work, we are expected to deeply analyze a signal data distribution from more perspectives, and further promote such unrolled network designs.

\bibliographystyle{IEEEtran}
\bibliography{references}

\end{document}